\documentclass[
 amsmath,amssymb,
 aps,
 prx,
 superscriptaddress,
 nofootinbib,
 twocolumn
]{revtex4-2}
\usepackage{graphicx}
\usepackage[normalem]{ulem}
\usepackage[colorlinks,linkcolor=blue,urlcolor=blue, citecolor=blue]{hyperref}
\usepackage{setspace}
\usepackage{multirow}
\usepackage{siunitx}
\usepackage[dvipsnames]{xcolor}
\usepackage{amsmath, amssymb, bm}
\usepackage{booktabs}
\usepackage{makecell}
\usepackage{mathtools}
\usepackage{adjustbox}
\usepackage{xcolor}
\usepackage{soul}
\usepackage{amsmath,amssymb,braket}
\usepackage{soul}

\usepackage{physics2}
\usephysicsmodule{ab}
\usephysicsmodule{op.legacy}
\usephysicsmodule{braket}
\usephysicsmodule{nabla.legacy}
\usepackage{derivative}
\newcommand{\mqty}[1]{\ab(\begin{matrix}#1\end{matrix})}
\newcommand{\dd}[1]{\odif{#1}\,}

\usepackage{xcolor}

\usepackage{comment}
\DeclareMathOperator*{\Maj}{\text{Maj}}
\DeclareMathOperator*{\wt}{\text{wt}}

\begin{document}
\title{Operational criteria for quantum advantage in latency-constrained nonlocal games}

\author{Changhao Li}
\email{changhao.li@nano-qt.com}
\thanks{equal contribution}
\affiliation{Nanofiber Quantum Technologies, Inc. (NanoQT), 313 Bryant Court, Palo Alto, CA 94301, USA}
\affiliation{Unitary Foundation, San Francisco, CA, USA}
\author{Seigo Kikura}
\email{seigo.kikura@nano-qt.com}
\thanks{equal contribution}
\affiliation{Nanofiber Quantum Technologies, Inc. (NanoQT), 1-22-3 Nishiwaseda, Shinjuku-ku, Tokyo 169-0051, Japan.}
\author{Akihisa Goban}
\email{akihisa.goban@nano-qt.com}
\affiliation{Nanofiber Quantum Technologies, Inc. (NanoQT), 1-22-3 Nishiwaseda, Shinjuku-ku, Tokyo 169-0051, Japan.}
\author{Hayata Yamasaki}
\email{hayata.yamasaki@nano-qt.com}
\affiliation{Nanofiber Quantum Technologies, Inc. (NanoQT), 1-22-3 Nishiwaseda, Shinjuku-ku, Tokyo 169-0051, Japan.}
\affiliation{Department of Computer Science, Graduate School of Information Science and Technology, The University of Tokyo, 7-3-1 Hongo, Bunkyo-ku, Tokyo 113-8656, Japan}
\author{Shinichi Sunami}
\email{shinichi.sunami@nano-qt.com}
\affiliation{Nanofiber Quantum Technologies, Inc. (NanoQT), 1-22-3 Nishiwaseda, Shinjuku-ku, Tokyo 169-0051, Japan.}
\affiliation{Clarendon Laboratory, University of Oxford, Oxford OX1 3PU, United Kingdom}

\begin{abstract}

Remote entanglement enables coordinated decision making without communication and produces correlations beyond those achievable by any classical strategy, representing a practical quantum advantage in time-critical distributed decision-making problems.
However, existing analyses of quantum-classical gaps in such latency-constrained tacit coordination (LCTC) have focused on idealized models that neglect the finite stationary window of the LCTC, finite operation times, and limited entanglement generation rates, leaving fundamental constraints unaccounted for.
In this work, we develop a comprehensive framework to quantitatively analyze quantum advantage in LCTC that explicitly incorporates finite-duration and finite-rate operations, as well as generalized utility structures with a limited stationary window.
These advances are made possible by adapting statistical certification methods for nonlocal games to the decision-making scenarios of LCTC, identifying operational criteria that must be satisfied by the hardware implementations to realize quantum advantage with sufficient statistical significance.
To meet the stringent criteria, we propose time-multiplexed, event-ready operations of cavity-assisted trapped-atom quantum network nodes that provide a continuous stream of entangled qubit pairs, with decision latencies of a microsecond and decision rates of $8\times 10^3~\text{s}^{-1}$ per channel for a representative metropolitan-scale $50$-km fiber network to keep up with the fast-changing environment, such as financial markets and electric grid networks.
These results bridge the gap between the theoretical notions of the quantum-classical gap in nonlocal games and concrete implementations that meet the stringent operational criteria for achieving robust quantum advantage in realistic coordination tasks.

\end{abstract}
\maketitle

\section{Introduction}

Nonlocality is one of the most striking features of quantum mechanics, which has been tested by the violation of Bell inequalities~\cite{Bell1964, Horodecki2009, Brunner2014RMP} using various experimental means~\cite{Aspect1981, Rowe2001, Hensen2015Nature, Rauch2018, Nadlinger2022, Storz2023}.
Among the many manifestations of nonlocal correlations, quantum advantage in \emph{nonlocal games} is among the most intuitive~\cite{Buhrman2010RMP, Cleve2004, Brassard2005}, where multiple parties try to perform certain coordinated tasks without direct communication allowed after the start of the game, while a referee scores the realized coordination.
In such a game, classical strategies utilize preshared randomness and agreed strategies, while quantum strategies are allowed to supplement their decision-making with a measurement of preshared entangled quantum states.
Quantum advantage arises from the finite gap between the maximal score attained by any classical strategy and the value obtained by a quantum strategy.
Such an advantage is reported across various game settings, such as the two-party Clauser-Horne-Shimony-Holt (CHSH) game~\cite{Clauser1969, Aspect1982}, multiparty games~\cite{Pan2000}, the magic-square game~\cite{Xu2022PRL}, and the odd-cycle game~\cite{Drmota2025}. 
Remarkably, the quantum advantage in nonlocal games does not rely on complexity-theoretic assumptions, nor does it require asymptotic scaling of the problem sizes for quantum advantage, and it can be realized without the substantial overhead of fault-tolerant quantum computing~\cite{gidney2025factor,Zhou2025}, making it a realistic target for near-term quantum devices.

The practical application of such a nontrivial behavior of quantum entanglement has long been explored in various fields, such as device-independent quantum key distribution~\cite{Zapatero2023npjQI}, random number generation~\cite{Pironio2010}, and self-testing of quantum devices~\cite{Supic2020}.
Recently, Ding and Jiang~\cite{2407.21723} proposed that the quantum advantage in nonlocal games can be directly exploited in distributed coordination tasks under latency constraints, which we denote as latency-constrained tacit coordination (LCTC) tasks, where the need for no-signaling coordination arises naturally due to the vast separation of required coordinated decision-making timescales and communication latencies.
They provided multi-venue high-frequency trading (HFT) as a motivating example, where trading decision making must be made on the order of microseconds while inter-venue communication latencies are orders of magnitude longer, for example, taking over \qty{100}{\us} to directly communicate between two major trading venues in New York.
Similar application examples have been proposed recently, such as network load balancing~\cite{Arun2025} and routing optimization~\cite{daSilva2025}, and a much wider variety of settings is expected to be identified and analyzed.

\begin{figure}[t]
    \centering
    \includegraphics[width=1.01\linewidth]{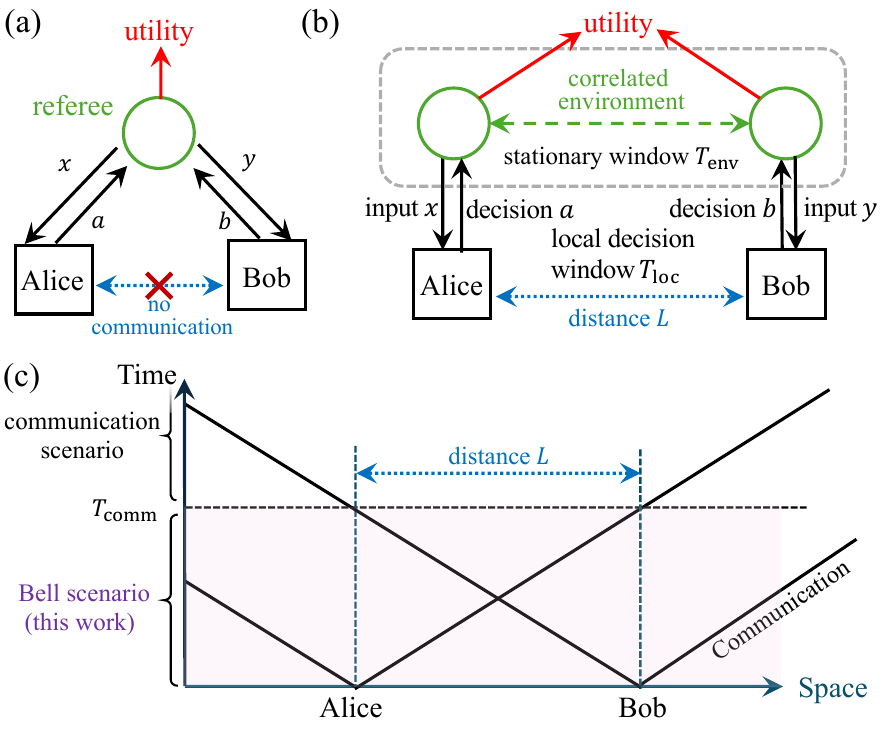}
    \caption{\textbf{Latency-constrained tacit coordination.}
    \textbf{(a)} Referee-based nonlocal game: inputs $(x,y)$ are distributed to distant parties who output $(a,b)$ and are scored (utility).
    \textbf{(b)} 
    LCTC task: inputs are local signals from a correlated environment located at spatial distance $L$, and the realized utility is accumulated over a duration $T_{\mathrm{env}}$ within which a utility function and an input distribution remain stationary. 
    Upon receiving signals $(x,y)$, decisions $(a,b)$ are made within a local decision time window $T_\mathrm{loc}$.
    \textbf{(c)} Spacetime structure for a round of LCTC task. For separation $L$ with required communication time $T_{\mathrm{comm}}$, the LC condition $T_{\mathrm{loc}}<T_{\mathrm{comm}}$ enforces a no-signaling (Bell) scenario, whereas $T_{\mathrm{loc}}>T_{\mathrm{comm}}$ would allow classical messaging within the decision window.
    While the illustrated light cones (black solid lines) resemble that of relativistic communication constraints, in realistic LCTC tasks, $T_\mathrm{comm}$ may be determined by more practical communication latencies that are slower than those determined solely by the speed of light in vacuum.
    }
    \label{fig:lc_tacit_coordination}
\end{figure}

One can view the setting of LCTC as a conceptual extension of the conventional nonlocal games (Fig.~\ref{fig:lc_tacit_coordination}).
In the conventional nonlocal games [Fig.~\ref{fig:lc_tacit_coordination}(a)], a \emph{referee} sends inputs to distant \emph{parties}, Alice and Bob, who return the outputs to the referee for scoring (\emph{utility}).
Here, the referee follows a fixed, predetermined input distribution and scoring (utility) rules.
In contrast, in LCTC, the nature of the coordination tasks dramatically changes [Fig.~\ref{fig:lc_tacit_coordination}(b, c)].
Here, there is no fixed referee, and the parties interact with their local \emph{environment}, which provides the parties with \emph{inputs} upon which they make their \emph{decisions} without direct communication.
The objective of the game is to perform coordinated decision making so that the parties exploit the inherent correlations between the two environments, maximizing their combined utility beyond what is possible with classical strategies.
Here, the no-signaling condition is imposed by the requirement for the parties to promptly make their decisions based on local inputs, faster than direct communication (\emph{Bell scenario}).
In the case of the HFT example, the environments are the two distant trading venues, which are usually strongly correlated, and the parties perform coordinated trading to exploit the inherent inter-venue correlation while maintaining a rapid response to local signals.

Such considerations highlight fundamental constraints that are absent in the conventional nonlocal game setting assumed for the analysis in Refs.~\cite{2407.21723,Arun2025,daSilva2025}, raising the question of whether quantum strategies can indeed provide a genuine quantum advantage in LCTC tasks;
the standard formulation of nonlocal games assumes an idealized setting of unbounded repetition under a fixed and symmetric game structure, whereas realistic LCTC tasks are constrained to a finite environmental stationary window $T_\text{env}$ within which the input distribution and utility structure remain valid, and beyond which the game itself changes. 
Furthermore, each round of quantum-assisted coordination incurs a finite duration for qubit measurement, and shared entanglement is generated only at a finite rate with imperfect fidelity. 
Therefore, the relevant question is not merely whether a positive quantum-classical gap exists in expectation, but whether a quantum strategy can yield a statistically certifiable advantage over any classical strategy within the limited number of rounds permitted by the environment. 
This question demands a unified treatment of the game structure, the certification statistics, and the performance of remote entanglement generation that supports the execution of a quantum strategy.

In this work, we establish \emph{operational criteria} to achieve robust quantum advantage in LCTC tasks, systematically incorporating key factors such as system asymmetry, noisy entanglement, finite-rate entanglement supply, and a finite environmental stationary window. 
More concretely, we first generalize the framework of LCTC in Ref.~\cite{2407.21723} to asymmetric system configurations and develop a concrete multiparty formulation, and analyze their noise tolerance; we further adapt the statistical certification methods developed for the tests of the local hidden-variable (LHV) model~\cite{Zhang2011PRA, Elkouss2016_pvalue, Arajo2020Quantum} to the time-critical nonlocal games played in LCTC tasks.
As an example, this allows us to translate the attainable rates of the game plays, i.e., the number of entangled qubit pairs supplied to the parties to enable the execution of quantum strategies in a fixed time window, into the statistical significance of the quantum advantage in LCTC tasks played under a limited stationary window of the environment, allowing us to map out the operational criteria to achieve a robust quantum advantage in realistic LCTC tasks.

To meet such a stringent requirement, we further propose a hardware implementation based on time-multiplexed, continuously operating quantum network nodes using trapped-atom qubits and optical cavities~\cite{Reiserer2015RMP}. 
This platform is a natural fit for LCTC because it combines the demonstrated capabilities of scalable, long-lived atomic qubit registers, and their high-fidelity control~\cite{Bluvstein2024,Manetsch2025Nature,Henriet2020Quantum} with cavity-assisted optical interfaces that can support telecom-band high-rate and high-fidelity remote entanglement generation~\cite{Covey2019,Li2024PRXQ,Sunami2025PRXQ,Sinclair2025PRR} and fast state readout~\cite{Deist2022PRL,Wang2025PRL,Grinkemeyer2025}.
We translate experimentally relevant parameters into system-level figures of merit, including entanglement generation rate, remote entanglement fidelity, measurement speed, and memory depth, which allow us to show that near-term implementation can satisfy the operational criteria for demonstrating robust quantum advantage in LCTC tasks requiring high decision rates.

These results are key to resolving the gap between idealized models of LCTC in the previous literature and a more realistic requirement for the remote  coordination tasks, allowing for a principled design of quantum strategies and a systematic analysis of the expected quantum advantage.
The refined conceptual model of the LCTC will also serve as a basis for extending the theoretical framework of LCTC.
We believe such an investigation will be further motivated by our realistic proposal for implementing quantum strategies for LCTC tasks that satisfy the stringent rate, fidelity, and measurement capability requirements; 
this architecture is expected to be applicable not only to LCTC tasks but also to other applications, for example, device-independent quantum key distribution~\cite{Zapatero2023npjQI}, quantum tokens~\cite{Horodecki_2020,Bozzio2025}, demonstration of energy-consumption quantum advantage~\cite{PRXEnergy.4.023008}, and the exploitation of quantum advantage in communication complexity~\cite{Buhrman2010RMP, LeGall2025, Tang2023BQC}, which will benefit from the high-rate and continuous generation of memory-based remote entanglement.

This paper is organized as follows. 
Section~\ref{sec:problem_statement} provides a formal introduction to nonlocal games and LCTC, reviews the existing statistical analysis methods of Bell test experiments, and includes a summary of representative applications of LCTC to motivate the relevant parameters analyzed in this work.
In Sec.~\ref{sec:operational_criteria}, we formalize a generalized LCTC task and identify the operational criteria to attain a statistically significant quantum advantage in LCTC under a limited time window and realistic noise on the shared entanglement and measurement operations. 
To meet the operational criteria, in Sec.~\ref{sec:multiplex}, we present a general architecture for time-multiplexed, event-ready quantum network nodes to continuously execute the quantum strategies of LCTC at high rates and quantify the required hardware performance.
In Sec.~\ref{sec:resource_estimation}, we propose a concrete instantiation of this network architecture with a cavity-assisted neutral-atom quantum network node, which simultaneously achieves high-rate remote entanglement generation and fast qubit measurement to support the execution of quantum strategies.
Section~\ref{sec:multipartite} extends the analysis to multiparty coordination using multi-qubit entanglement and proposes a neutral-atom implementation of high-rate multipartite entanglement generation.
Finally, Sec.~\ref{sec:discussions_outlook} concludes with a discussion of broader implications and experimental outlook.

\section{Nonlocal games and latency-constrained tacit coordination}\label{sec:problem_statement}
In this section, we establish the theoretical foundations for analyzing quantum advantage in LCTC tasks. We begin in Sec.~\ref{sec:non-local_games} by reviewing the formulation of nonlocal games, which provides the mathematical framework for characterizing the quantum and classical strategies that spatially separated parties employ. In Sec.~\ref{sec:lc_tacit_coord}, we specialize this framework to LCTC, where the no-signaling condition is enforced physically by communication delays relative to a strict local decision window. 
We then discuss the statistical certification of nonlocal games in Sec.~\ref{sec:finite_stats}, on which some of our operational criteria are based. 
Finally, Sec.~\ref{sec:application_examples} outlines concrete application scenarios, such as high-frequency trading and power grid management, illustrating several concrete examples of latency constraints and utility structures.

\subsection{Nonlocal games and quantum-classical gap} \label{sec:non-local_games}

In a standard bipartite \textit{nonlocal game}, a referee samples the inputs, i.e., $x,y \in \{0,\cdots, m-1\}$, where $m \in \mathbb{N}_+ = \{1,2,\cdots \}$, according to an input distribution $P(x,y)$ and respectively distributes $x, y$ to two spatially separated parties, Alice and Bob [see Fig.~\ref{fig:lc_tacit_coordination}(a)]; each input is unknown to the other party.
Upon receiving their respective inputs, Alice and Bob produce outputs 
$a, b \in \{0,\cdots,\Delta-1\} \,(\Delta \in \mathbb{N}_+)$ with their strategy characterized by the set of $\Delta^2 m^2$ conditional probabilities $\textbf{P} = \{P(a,b\mid x,y)\}$, called \emph{behavior}.
Here, the behavior is required to
 satisfy the no-signaling constraints such as $\sum_{b}P(a,b\mid x,y) = \sum_b P(a,b\mid x,y^\prime)$ for any $y$ and $y^\prime$~\cite{Brunner2014RMP}.
The two parties then send their output to the referee, and the referee evaluates the performance according to a utility function $u(a,b\mid x,y)\in{\mathbb{R}}$. 
The expected utility achieved by a given behavior is
\begin{equation}   \label{eq:expected_utility}
    \omega
    = \sum_{x,y} P(x,y)\sum_{a,b} u(a,b\mid x,y)\, P(a,b\mid x,y).
\end{equation}

We distinguish classical and quantum strategies by the behavior $\textbf{P}$ they can realize~\cite{Brunner2014RMP,Buhrman2010RMP}. 
We first define the set $\mathcal{L}$ of local behaviors such that the elements of a behavior $\textbf{P} \in \mathcal{L}$ satisfy
\begin{equation}
    P(a,b\mid x,y) = \int \dd{\lambda}P(\lambda)P_A(a\mid x,\lambda)P_B(b\mid y,\lambda),
\end{equation}
where $\lambda$ represents shared randomness, and $P_A(a\mid x,\lambda)$ and $P_B(b\mid y,\lambda)$ represent conditional probability distributions for Alice and Bob, respectively.
We call an operation that realizes a local behavior a classical strategy~\cite{2407.21723}, and maximizing the expected utility over all classical strategies defines the classical value $\omega_{\mathrm C}$.
On the contrary, if the elements of a behavior consist of a shared bipartite quantum state $\rho$ and local measurements described by positive operator valued measures (POVMs), $\{\Pi_{a\mid x}\}_a$ and $\{\Pi_{b\mid y}\}_b$, as
\begin{equation} \label{eq:quantum_strategy}
    P(a,b\mid x,y) = \Tr\ab[\rho(\Pi_{a\mid x} \otimes \Pi_{b\mid y})],
\end{equation}
the behavior belongs to the set $\mathcal{Q}$ of quantum behaviors.
To explicitly incorporate noisy entanglement, we define a quantum strategy as a tuple $(\rho, \{\Pi_{a\mid x}\}_a, \{\Pi_{b\mid y}\}_b)$, similar to Ref.~\cite{2407.21723}, and the set of quantum behaviors realized by the set of implementable quantum strategies $\{(\rho, \{\Pi_{a\mid x}\}_a, \{\Pi_{b\mid y}\}_b)\}$ as $\mathcal{\bar{Q}} (\subseteq \mathcal{Q})$.
The quantum value $\omega_\text{Q}$ is then defined as the maximized expected utility over any behaviors in the set $\mathcal{\bar{Q}}$.
Finally, a quantum advantage corresponds to a positive quantum-classical gap in expected utility
\begin{equation} \label{eq:Delta_U_def}
\Delta \omega = \omega_\text{Q} - \omega_{\mathrm C}.
\end{equation}
Throughout this work, $\Delta \omega$ serves as the key metric for quantifying quantum advantage, capturing the maximal performance gain achievable using quantum resources relative to all classical strategies.

One of the well-studied classes of nonlocal games is an XOR game: each player has binary inputs $x,y\in\{0,1\}$ and outputs $a,b \in \{0,1\}$, and the utility function is in the form of
\begin{align}
    u_\mathrm{XOR}(a,b\mid x,y)= u(a\oplus b \mid x, y),
\end{align}
where $a \oplus b = a+b ~ \text{mod}~2$~\cite{Brunner2014RMP}.
In an XOR game, a strategy can be characterized by a correlator
\begin{equation}
    E_{x,y} =  \sum_{a,b}(-1)^{a\oplus b} P(a,b\mid x,y),
\end{equation}
and the expected utility is rewritten by~\cite{Brunner2014RMP}
\begin{equation}  \label{eq:omega_XOR}
    \omega = \frac{1}{2}\sum_{x,y}P(x,y) \sum_{o\in \{0,1\}} u(o\mid x,y) [1+(-1)^o E_{x,y}].
\end{equation}
This means that a strategy aims to maximize $\sum_{x,y} M_{x,y}E_{x,y}$ with
\begin{equation} \label{eq:M_XOR}
    M_{x,y} = P(x,y) \sum_{o\in \{0,1\}} (-1)^o u(o\mid x,y),
\end{equation}
which is only characterized by the game setting.
For a classical strategy, the maximized value can be achieved by a deterministic strategy, leading to~\cite{Brunner2014RMP,Acin2006, Vrtesi2008PRA}
\begin{equation} \label{eq:C(M)_def}
    \max_{\textbf{P}\in \mathcal{L}} \sum_{x,y} M_{x,y}E_{x,y} = \max_{n_a , n_b \in\{\pm{1}\}} \sum_{x,y} M_{x,y}n_an_b \eqqcolon C(M).
\end{equation} 
In contrast, for a quantum strategy, Alice and Bob each choose $\pm 1$-valued observables $A_x$ and ${B_y}$, leading to projective measurements
\begin{equation}
    \Pi_{a\mid x} = \frac{\mathbb{I}_A +(-1)^{a} A_x}{2}, \quad  \Pi_{b\mid y} = \frac{\mathbb{I}_B +(-1)^{b} B_y}{2},
\end{equation}
where $\mathbb{I}_{A(B)}$ represents an identity operator in a Hilbert space for Alice (Bob).
This results in
\begin{equation} \label{eq:E^Q_general}
    E_{x,y} = \Tr[\rho (A_x \otimes B_y)],
\end{equation}
and aims to maximize $\sum_{x,y}M_{x,y}E_{x,y}$ over $\textbf{P}\in \mathcal{\bar{Q}}$.
We finally note that an XOR game in which the utility function is normalized as $\sum_{o \in\{0,1\}} u(o\mid x,y) = 1$ for any inputs $(x,y)$ simplifies the expected utility as
\begin{equation} \label{eq:omega_XOR_game_simplified}
    \omega = \frac{1}{2}\ab(1 + \sum_{x,y} M_{x,y}E_{x,y}).
\end{equation}
 
As a canonical example, we consider the Clauser–Horne–Shimony–Holt (CHSH) game~\cite{Clauser1969,Cleve2004} with binary inputs $x,y\in\{0,1\}$ drawn uniformly $[P(x,y) =1/4]$. The utility indicating the win or lose of the game is defined as
\begin{equation} \label{eq:chsh_predicate}
    u_\text{CHSH}(a,b\mid x,y)=
    \begin{cases}
    1, & a \oplus b = x y,\\
    0, & \text{otherwise},
\end{cases}
\end{equation}
leading to the expected utility in Eq.~\eqref{eq:omega_XOR_game_simplified} with $M_{x,y} = (-1)^{xy}/4$.
In this case, any classical strategy satisfies $\sum_{x,y}M_{x,y}E_{x,y} \leq 1/2$ through the CHSH inequality, leading to $\omega_\text{C} = 0.75$.
In contrast, a quantum strategy can achieve $\max_{\textbf{P}\in \mathcal{Q}} \sum_{x,y}M_{x,y}E_{x,y} = 1/\sqrt{2}$ by measuring a singlet state
\begin{align}
\label{eq:singlet}
    \ket{\Psi^-}=\frac{1}{\sqrt{2}}(\ket{01}-\ket{10}),
\end{align}
with appropriately chosen measurement basis~\cite{Brunner2014RMP}, resulting in the theoretically optimal quantum value $\omega_\text{Q} = (1+1/\sqrt{2})/2 > 0.85$.
This gap $\Delta \omega = (\sqrt{2}-1)/4$ provides a standard benchmark for how shared entanglement can increase achievable utility under no-signaling constraints.

\subsection{Latency-constrained tacit coordination}
\label{sec:lc_tacit_coord}

We now consider the LCTC task~\cite{2407.21723} as an extension of the nonlocal games, as illustrated in Fig.~\ref{fig:lc_tacit_coordination}(b), which introduces additional temporal constraints to be satisfied.
In particular, we introduce a finite local decision window $T_{\mathrm{loc}}$, which is the time available from the arrival of the local input ($x, y$) at each party to the time the corresponding decision  ($a, b$) must be made: Alice and Bob are required to produce decisions within $T_{\mathrm{loc}}$. 
Let $T_\mathrm{comm}$ denote the minimum time required for any classical messages to be communicated between parties, which is nonzero due to both relativistic and technological limitations [Fig.~\ref{fig:lc_tacit_coordination}(c)]. 
The LC condition is given by
\begin{equation}  \label{eq:lc_condition}
    T_\mathrm{loc}<T_\mathrm{comm}, 
\end{equation}
where any communication between Alice and Bob during the local decision window is prohibited. 
Under this condition, each party’s output can depend only on its local input and any resources established prior to the arrival of inputs. 
As a result, the effective strategies available in an LCTC task are precisely those described in Sec.~\ref{sec:non-local_games}: classical strategies based on shared randomness and quantum strategies based on shared entanglement and local measurements. 
Therefore, each round of an LCTC task corresponds to a single instance of the nonlocal game defined in Sec.~\ref{sec:non-local_games}, with performance quantified by the expected utility $\omega$.

\subsection{Finite-statistics effects and finite-round certified quantum advantage}\label{sec:finite_stats}

The quantum-classical gap discussed in Sec.~\ref{sec:non-local_games} is defined at the level of expectation values: for a fixed LC game, $\Delta \omega >0$ indicates a quantum advantage that is relevant in the limit of large statistics.
One might be tempted to believe that such a game could be repeated indefinitely and that the expected utility gap $\Delta \omega$ might be a sufficient measure of quantum advantage; however, many realistic LCTC tasks feature only a finite stationary window for the exploitable correlations.
Even if the correlations themselves may be mostly stationary, the resulting utility structure will likely drift over time.
This motivates the introduction of an additional parameter $T_\mathrm{env}$, which denotes the finite stationary window of the given utility structure [Fig.~\ref{fig:lc_tacit_coordination}(b)].
This will be further motivated by the concrete examples of LCTC, discussed in Sec.~\ref{sec:application_examples}.

In any practical setting with finite $T_\mathrm{env}$, the realizations of nonlocal games are limited to a finite number of rounds; 
even when $\Delta \omega>0$, a classical strategy may outperform the quantum strategy over a short run purely by chance. 
This motivates the adoption of certified advantage within finite statistics~\cite{Elkouss2016_pvalue, Arajo2020Quantum}, closely paralleling the statistical analysis used to reject LHV models, which we review in the following.

We first consider that the utility $u(a,b\mid x,y)$ represents the probability that the referee judges the parties to win, which reduces $\omega$ to winning probabilities~\cite{Arajo2020Quantum}. 
In this case, the $p$-value, i.e.,~the maximum probability of obtaining $v$ or more winning events out of $m$ rounds with LHV models, is given by~\cite{Elkouss2016_pvalue,Arajo2020Quantum}
\begin{equation}  \label{eq:pvalue_binomial}
    p(v,m) =\sum_{k=v}^{m}\binom{m}{k}\omega_\text{C}^{\,k}(1-\omega_\text{C})^{m-k}.
\end{equation}
In experiments aimed at testing LHV models, the obtained $v$ winning events among $m$ rounds indicate the rejection of LHV models with a significance level $\alpha = p(v,m)$~\cite{Rosenfeld2017}.
Here, we use this idea not primarily to make foundational claims about local realism, but rather to translate a target significance level $\alpha$ for quantum advantage in LCTC into a required number of game rounds, which can then be converted into a requirement on the sustainable game-play and entanglement rate in Sec.~\ref{sec:operational_criteria}.
To this end, we set $v$ as the expected number of wins $\lceil m\omega_\text{Q}\rceil$ with a set $\mathcal{\bar{Q}}$ of quantum strategies~\cite{Arajo2020Quantum}, and define the required number $n_\text{req}(\alpha)$ of total game rounds with a significance level $\alpha$ as 
\begin{equation} \label{eq:n(alpha)_def}
    n_\text{req}(\alpha) = \min\{m\in \mathbb{N}_+ \mid p(\lceil m\omega_\text{Q}\rceil, m) < \alpha \},
\end{equation}
where $\lceil\cdot\rceil$ is the ceiling function.
This means that by playing $n_\text{req}(\alpha)$ rounds, the probability that any classical strategy exceeds the quantum expectation is less than $\alpha$.
Thus, we must play nonlocal games for at least $n_\text{req}(\alpha)$ rounds, within the time window during which $P(x,y)$ and $u(a,b \mid x,y)$ remain fixed, that is, within a duration of $T_\text{env}$.
In other words, the required rate of game instance is
\begin{equation} \label{eq:R_dec(alpha)}
    R_\text{req}(\alpha) = \frac{n_\text{req}(\alpha)}{T_\text{env}}.
\end{equation}

For a generalized nonlocal game, where $u_\text{max} = \max_{a,b,x,y} u(a,b\mid x,y)$ and/or $u_\text{min} = \min_{a,b,x,y} u(a,b\mid x,y)$ may take values outside the range $[0,1]$, the utility $u(a,b\mid x,y)$ can be interpreted as the score rather than the win probability~\cite{Elkouss2016_pvalue}.
If the total score over $m$ game rounds is obtained by $c$, the $p$-value for the observed score is given as the maximum probability of obtaining a total score $C \geq c$ with classical strategies~\cite{Elkouss2016_pvalue};
\begin{equation}
    p(c, m) = \max_{\text{classical}}\Pr[C\ge c \mid \text{classical strategy}].
\end{equation}
This probability is bounded by~\cite{Elkouss2016_pvalue}
\begin{equation} \label{eq:p(c,n)_general_game}
    p(c, m) \leq e \ab[P_{m,\lfloor\mu\rfloor}(\mathbb{B}_{\xi})]^{1-\mu + \lfloor\mu\rfloor} \ab[P_{m,\lceil\mu\rceil}(\mathbb{B}_{\xi})]^{\mu - \lfloor\mu\rfloor},
\end{equation}
where $e$ is Napier's constant, $\lfloor\cdot\rfloor$ represents the floor function,
\begin{equation}
    \mu  = \frac{c - m u_\text{min}}{u_\text{max}-u_\text{min}},\quad \xi = \frac{\omega_\text{C} - u_\text{min}}{u_\text{max} - u_\text{min}},
\end{equation}
and we define
$P_{m,k}(\mathbb{B}_{\xi}) = \sum_{i=k}^m \binom{m}{i} {\xi}^{i}(1-{\xi})^{m-i}$.
Setting $c = \lceil m\omega_\text{Q}\rceil$ again gives the required number $n_\text{req}(\alpha)$ to satisfy Eq.~\eqref{eq:n(alpha)_def}. 
We finally note that since the results of the $p$-value in Eqs.~\eqref{eq:pvalue_binomial} and~\eqref{eq:p(c,n)_general_game} hold for an arbitrary number of parties~\cite{Elkouss2016_pvalue}, the proposed framework for finite-round certified quantum advantage can be employed for multiparty nonlocal games discussed in Sec.~\ref{sec:multipartite}.

\subsection{Application examples}\label{sec:application_examples}

In this section, we highlight three representative applications of the LCTC tasks discussed in the previous section, which have the latency constraint of Eq.~\eqref{eq:lc_condition} and the utility structure representing a finite advantage for coordinated actions.

\subsubsection{High-frequency trading}

As a concrete case study, we consider the application in high-frequency trading proposed by Ding and Jiang~\cite{2407.21723}. 
In modern electronic markets, HFT strategies seek to capture transient pricing discrepancies while managing risks by reacting to market data on the order of microseconds to tens of microseconds, where profitability depends on making decisions within a short duration during which a price discrepancy remains exploitable~\cite{Aquilina2021}.
LCTC naturally arises when one wishes to trade strongly correlated financial instruments across multiple geographically separated trading venues, for example, between the National Association of Securities Dealers Automated Quotations (NASDAQ) and the New York Stock Exchange (NYSE) separated by $\qty{56.3}{\km}$.
The LC condition arises because classical inter-venue signal propagation $T_\mathrm{comm}$ far exceeds the typical local decision window $T_\mathrm{loc}$ of several microseconds; even for the moderate case of $\qty{56.3}{\km}$ direct line-of-sight separation between NASDAQ and NYSE, $T_\mathrm{comm}$ is at least $\qty{188}{\us}$.
Here, $T_\mathrm{env}$ represents the time window during which the inter-venue correlation and the utility structure remain stationary.
While the exact time window $T_\mathrm{env}$ may vary widely depending on the specific instrument pairs being traded and overall market conditions, we adopt a representative timescale on the order of seconds, which is reported in recent studies as the time range during which a pre-trained statistical model of the price dynamics remains valid.
For example, Ref.~\cite{AtSahalia2022} reports that short-term future price movements of heavily traded stocks can be predicted from their recent trading history, with the predictability being strongest over the next few seconds and decaying rapidly thereafter; similar timescales for the decay of cross-asset predictability are reported in Refs.~\cite{Huth2014HFT,Brogaard2014HFT}.

As an exemplary formulation of such an LCTC task, we follow the hedging problem introduced in Ref.~\cite{2407.21723}.
Consider a market maker operating at two venues that trade correlated stocks. 
As a market maker, the trader conventionally places both a bid and an ask order in each venue; due to the sequential nature of their submission to the respective trading system, ask-first (\(0\)) or bid-first (\(1\)) submission ordering decisions ($a,b\in\{0,1\}$) result in slightly different economic return structures.
Let \(x,y\in\{0,1\}\) denote whether the respective server observes a local indicator that the usual correlation pattern has temporarily changed. In this toy model, the utility depends only on whether the two servers make the same decision ($o=a\oplus b=0$) or opposite decisions ($o=1$). 
Thus, up to a relabeling of the binary observations and actions, the task reduces to the XOR-game utility function introduced in Ref.~\cite{2407.21723}:
\begin{equation}     \label{eq:utility_func_HFT}
    u_{\mathrm{HFT}}(o \mid x,y)=
    \begin{cases}
        1, & (x,y)=(0,0),\ o=0,\\
        1-\beta, & (x,y)=(0,1),\ o=0,\\
        1-\beta, & (x,y)=(1,0),\ o=0,\\
        1, & (x,y)=(1,1),\ o=1,\\
        \beta, & (x,y)=(0,1),\ o=1,\\
        \beta, & (x,y)=(1,0),\ o=1,\\
        0, & \text{otherwise},
    \end{cases}
\end{equation}
where $\beta \in [0,1]$ represents a partial preference for specific coordinations in the case of only one of the servers observing the signal. 
Note that $\beta = 0$ reduces the function to $u_\text{CHSH}(a,b\mid x,y)$ in Eq.~\eqref{eq:chsh_predicate}.

\subsubsection{Power grid management}

Electric power grids maintain a reliable electricity supply by continuously taking appropriate stabilization measures at control stations distributed across the grid. 
One example of the most time-critical operations in power grid management is the 
fast distributed grid-response~\cite{NERC_guideline_ForcedOscillation,NERC_guideline_PMU}, where two distant controllers detect local alarm signals and must each choose one of the pre-approved response modes before detailed information can arrive through the classical network.
In such fast distributed grid-response settings, because the protective mode responses must be taken before the  transient power instability propagates through the grid, the effective local decision window can be on the order of milliseconds~\cite{NERC_guideline_PMU}, which specifies $T_\mathrm{loc}$ for this application. 
In contrast, the end-to-end communication latency $T_\mathrm{comm}$ between geographically separated controllers can be hundreds of milliseconds and even several seconds due not only to relativistic limits but also to preprocessing and other network overheads, such as cybersecurity measures and other potential network delays~\cite{NERC_guideline_PMU}.
Under such conditions, coordinated protective action must rely on local information and pre-established strategies, naturally mapping onto the LCTC. 
Finally, the reported timescales for grid disturbances, which determine $T_\mathrm{env}$, are typically on the order of minutes~\cite{NERC_guideline_ForcedOscillation}.

As a representative benchmark, let \(x,y\in\{0,1\}\) denote whether the respective controller observes a local signal that indicates a possible fault, and let \(a,b\in\{0,1\}\) denote the two types of protective operating modes that each controller can decide on, for example, a rapid local shedding (0) or a less drastic curtailment mode or a hold mode (1)~\cite{LoadControlReview2022}.
The desired compatibility relation is \(a \oplus b = xy\): when both controllers observe possible faults, they should diversify to avoid redundant, high-cost actions; whereas in all other cases, they should select the same mode to maintain a consistent grid response.
The corresponding utility function is the same as Eq.~\eqref{eq:utility_func_HFT}.

\subsubsection{Network load balancing}

In data centers and geographically distributed cloud systems, data traffic must be dynamically assigned by multiple \emph{load balancers} to avoid congestion and maintain throughput; 
these must operate in a coordinated manner, often using only locally observed indicators to optimize total information throughput and avoid congestion. 
These locally available indicators include link utilization and congestion signals, and must be acted upon before detailed diagnostic information from distant nodes can arrive~\cite{Arun2025,Hu2022LoadBalancing}.

Relevant timescales vary widely depending on the geographical distribution of the network and the detailed choice of communication protocols; however, latency-constrained tasks commonly arise.
Local routing decisions are often made within a microsecond or several microseconds, while collecting congestion diagnostic information often takes tens of microseconds, even within one data center. 
The communication latency is much longer for geographically separated network nodes, related to the so-called \emph{round-trip time} of network protocols~\cite{Alizadeh2014}, as often reported in the literature~\cite{OrbWeaver2022,Alizadeh2014,Vasudevan2009}; in such cases, coordination must rely solely on local indicators and pre-established strategies, a setting fully compatible with LCTC tasks~\cite{Arun2025}.
The duration of frequent network congestion varies widely, with short congestion lasting from tens of milliseconds to longer, more severe events persisting for extended durations, such as those lasting over tens of seconds; thus, we consider the stationary window of $T_\mathrm{env} = \qty{10}{\ms}$--$\qty{10}{\s}$, with typically increasing severity and decreasing frequency for longer duration congestion events.

It is convenient to describe a representative load-balancing example in terms of the parity \(o=a\oplus b\) of two pre-approved coordination modes. 
Let \(x,y\in\{0,1\}\) denote local congestion indicators at the two sites, and let \(a,b\in\{0,1\}\) denote the selected pre-agreed coordination mode.
Concretely, a coordination mode should be understood as one of two admissible rules specifying how incoming traffic is assigned among available paths based on the local congestion indicator.
For the utility function, here we show the one consistent with what was recently presented in Ref.~\cite{Arun2025}, with modifications made for a more widely applicable case of asymmetric load. The application rewards opposite modes when both sites indicate congestion, while otherwise matching modes are preferred. This is represented by a generalized XOR utility function, namely
\begin{equation}
u_{\mathrm{LB}}(o \mid x,y)=
\begin{cases}
1, & (x,y)=(0,0),\ o=0,\\
1-\beta_1, & (x,y)=(0,1),\ o=0,\\
1-\beta_2, & (x,y)=(1,0),\ o=0,\\
1, & (x,y)=(1,1),\ o=1,\\
\beta_1, & (x,y)=(0,1),\ o=1,\\
\beta_2, & (x,y)=(1,0),\ o=1,\\
0, & \text{otherwise},
\end{cases}
\label{eq:utility_lb}
\end{equation}
where $\beta_1, \beta_2 \in [0,1]$.
The case of \(\beta_1\neq\beta_2\) captures coordinated load balancing tasks with system asymmetry, such as unequal traffic volume, priority, or traffic assignment costs in realistic coordination tasks, highlighting the relevance of the generalized utility structure \(u(a,b\mid x,y)\neq u(a,b\mid y,x)\) where the mixed-input cases, \((1,0)\) and \((0,1)\), can carry different costs. 
This motivates the asymmetric utility weights and correlated inputs introduced in Sec.~\ref{sec:extended_protocols}.

Table~\ref{tab:timescales} summarizes the representative timescales of the three application examples.
In each case, the LC condition $T_\mathrm{loc} < T_\mathrm{comm}$ is satisfied, while $T_\mathrm{env}$ sets the required decision rate [see Eq.~\eqref{eq:R_dec(alpha)} and Sec.~\ref{sec:operational_criteria}].

\begin{table}
    \caption{Representative timescale orders for the LCTC applications discussed in Sec.~\ref{sec:application_examples}.}
    \small
    \begin{ruledtabular}
        \begin{tabular}{lccc}
            Application & $T_\mathrm{loc}$ & $T_\mathrm{comm}$ & $T_\mathrm{env}$ \\
            \hline
            High-frequency trading & 1--\qty{10}{\us} & $>$ \qty{100}{\us} & 1--\qty{10}{\s} \\
            Power grid management & 1--\qty{10}{\ms} & \qty{100}{\ms}--\qty{1}{\s} & $>$ \qty{1}{\minute} \\
            Network load balancing & 1--\qty{10}{\us} & \qty{10}{\us}--\qty{10}{\ms} & \qty{10}{\ms}--\qty{10}{\s} \\
        \end{tabular}
    \end{ruledtabular}
    \label{tab:timescales}
\end{table}

\section{Operational criteria for quantum advantage} \label{sec:operational_criteria}

\begin{figure*}[t]
    \centering
\includegraphics[width=1\linewidth]{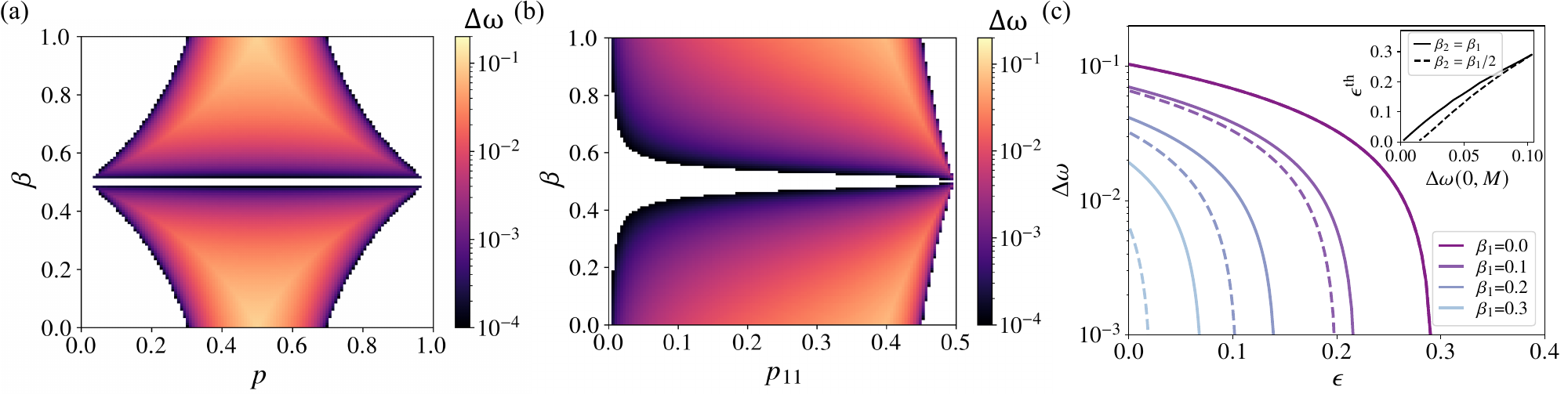}
    \caption{\textbf{Quantum-classical gaps in generalized LCTC tasks.}
    \textbf{(a)} The quantum-classical gap $\Delta \omega$ for independent and identically distributed Bernoulli inputs with bias $p$ and symmetric utility parameter $\beta_1=\beta_2=\beta$, assuming an ideal quantum strategy ($\epsilon=0$), where the combined infidelity $\epsilon$ accounts for infidelities of both states and measurements; color shows $\Delta \omega$ on a logarithmic scale (shown down to $\Delta \omega<10^{-4}$).
    \textbf{(b)} The gap $\Delta \omega$ for representative correlated input distributions, $p_{11} = P(1,1) = 2P(0,1) = 2P(1,0)$ for $\beta_1=\beta_2=\beta$.
    \textbf{(c)} The gap $\Delta \omega$ versus the combined infidelity $\epsilon$ for several $\beta$ at uniform inputs [$P(x,y)=1/4$]; 
    solid lines are for $\beta_2=\beta_1$, while dashed lines are for $\beta_2 = \beta_1/2$.
    Inset shows the threshold $\epsilon^\text{th}$ required for $\Delta \omega>0$ versus the ideal gap $\Delta \omega(\epsilon=0, M)$, for $\beta_2=\beta_1$ (solid) and $\beta_2=\beta_1/2$ (dashed) for range $\beta_1 \in [0,0.5]$.}
    \label{fig:extended_protocols}
\end{figure*}

Based on the discussions in the previous section, we derive the \emph{operational criteria} for achieving certified quantum advantage in LCTC tasks in the presence of a general utility structure under a limited stationary time window, a finite local decision window, and LCTC task trial rates.
In Sec.~\ref{sec:extended_protocols}, we introduce a generalized LCTC task involving weighted utilities, general input distributions and noisy operations, on which the subsequent criteria analysis is based.
In Sec.~\ref{sec:requirement_on_fidelity}, we then show the \emph{fidelity criterion}: a fidelity threshold of a feasible quantum strategy to achieve a positive gap $\Delta \omega >0$.
We also show the dependence of the achievable gap on the fidelity.
In Sec.~\ref{sec:rate_requirement}, we qualitatively analyze  the \emph{rate criterion}: the required rate of LCTC task execution to achieve statistically significant quantum advantage.
Finally, in Sec.~\ref{sec:decision_criteria}, we discuss the \emph{decision criterion}, the requirement for the speed of decision making.
    
\subsection{Setup: generalized LCTC tasks under noisy operations} \label{sec:extended_protocols}

To present general operational criteria applicable to a wide range of LCTC tasks, we first discuss generalized LCTC tasks in the presence of an asymmetric utility structure and generalized input distributions, as well as the quantitative treatment of the effect of noisy entanglement generation and noisy measurement.
More concretely, we extend the LCTC tasks in Ref.~\cite{2407.21723}, where each input was independently sampled from the same distribution: $P(x,y) = P(x) P(y)$ and $P(x) = P(y)$ where $P(x) = \sum_{y} P(x,y)$ and $P(y) = \sum_{x} P(x,y)$~\cite{2407.21723}, to two regimes relevant for deployment; asymmetric utility weights ($\beta_1,\beta_2$) and a correlated input distribution $P(x,y) \neq P(x) P(y)$.
These extensions are required to apply a nonlocal game to a general class of LCTC tasks, as discussed in Sec.~\ref{sec:application_examples}.
For instance, the parties' inputs may be correlated, or the utility does not satisfy $u(a,b\mid x,y) = u(a,b\mid y,x)$.
Such features arise in settings ranging from multi-venue high-frequency trading with asymmetric sensitivity to shared market sentiment to networked systems with non-uniform congestion signals and asymmetric priority for the use of limited data throughput.

In this generalized model, we focus on binary inputs $x,y\in\{0,1\}$ drawn from $P(x,y)$ and binary actions $a,b\in\{0,1\}$, which constitute the minimal setting in which quantum strategies can yield an advantage in LCTC tasks.
As an example, we adopt the utility function of Eq.~\eqref{eq:utility_lb}, equivalently rewritten as
\begin{equation} \label{eq:utility_AIC}
    u_\mathrm{LB}(o\mid x,y) 
    = \begin{cases}
        (1-\beta_{x+1})^{x\oplus y}, & o = xy,\\
        (x\oplus y)\beta_{x+1}, & \text{otherwise},
    \end{cases}
\end{equation}
where the parameters $\beta_1,\beta_2 \in [0,1]$ represent the penalty for mismatched actions on different inputs $x\neq y$, allowing heterogeneous tolerance for coordination failure across the two parties. 
Note that $P(x,y) = 1/4$ and $\beta_1=\beta_2=0$ reduce to the canonical CHSH game in Eq.~\eqref{eq:chsh_predicate}.
From the fact that $\sum_{o=\{0,1\}}u_\mathrm{LB}(o\mid x,y) = 1$, the expected utility is given in the form of Eq.~\eqref{eq:omega_XOR_game_simplified}, where the $2\times 2$ matrix $M = (M_{x,y})$ is
\begin{equation}
    M = \mqty{P(0,0) & P(0,1)(1-2\beta_1) \\ P(1,0) (1-2\beta_2) & P(1,1)}.
\end{equation}

We now turn to the effect of noise in the entangled states and measurement operations.
First, consider the general two-qubit state $\rho$ with imperfect entanglement, $1-\epsilon_\text{s} = \bra{\Psi^{-}}\rho\ket{\Psi^{-}} < 1$, where $\epsilon_\text{s}$ is the infidelity, and $\ket{\Psi^{-}}$ is the singlet state defined in Eq.~\eqref{eq:singlet}.
While there is a wide variety of noise models to be considered, here we focus on states of a Werner form~\cite{Almeida2007}
\begin{equation} \label{eq:werner}
    \begin{aligned}
        \rho(\epsilon_\text{s}) =& (1-\epsilon_\text{s}) \ketbra{\Psi^{-}}{\Psi^{-}} + \frac{\epsilon_\text{s}}{3}\ab(\mathbb{I}^{\otimes 2} - \ketbra{\Psi^{-}}{\Psi^{-}}) \\
        =& \ab(1- \frac{4\epsilon_\text{s}}{3}) \ketbra{\Psi^{-}}{\Psi^{-}} + \frac{\epsilon_\text{s}}{3}\mathbb{I}^{\otimes 2},
    \end{aligned}
\end{equation}
where $\mathbb{I}$ represents the identity operator on a qubit.
This form is widely relevant since one can effectively map other noisy entanglements to this form by randomly applying one of the bilateral single-qubit rotations~\cite{Werner1989, Bennett1996mixedstate}.
For any single-qubit measurement in an orthonormal basis $\{\ket{\psi_0},\ket{\psi_1}\}$ of $\mathbb{C}^2$, the measurement is represented as a traceless observable $(+1)\ketbra{\psi_0}{\psi_0}+(-1)\ketbra{\psi_1}{\psi_1}$ with $\pm 1$ eigenvalues.
Two parties choose their observables as $A_x = \sum_{i} (\hat{a}_x)_i \sigma_i$ and $B_y = \sum_{i} (\hat{b}_y)_i \sigma_i$, where $\sigma_1=X, \sigma_2=Y$ and $\sigma_3=Z$ are the Pauli operators
\begin{equation}
    X = \mqty{0 & 1\\ 1& 0}, \quad  Y = \mqty{0 & -i \\ i& 0},\quad Z = \mqty{1 & 0\\ 0&-1}
\end{equation}
in the qubit basis, and $\hat{a}_x, \hat{b}_y \in \mathbb{R}^3$ are unit vectors that characterize the measurement axes on the Bloch sphere.
Here, we consider the measurement result to be flipped by a probability $\epsilon_\text{meas}$, representing measurement infidelity.
This effect can be modeled by transforming the observables as,
\begin{equation} \label{eq:A_x(epsilon)}
    \begin{aligned}
        A_x(\epsilon_\text{meas}) =& (1-\epsilon_\text{meas})A_x + \epsilon_\text{meas} (-A_x) \\
        =& (1-2\epsilon_\text{meas})A_x,
    \end{aligned}
\end{equation}
and analogously for $B_y(\epsilon_\text{meas})$.
As a result, the correlator in Eq.~\eqref{eq:E^Q_general} becomes~\cite{Vrtesi2008PRA, Acin2006}
\begin{equation}\label{eq:meas_infid}
    \begin{aligned}
        E_{x,y}(\epsilon_\text{s},\epsilon_\text{meas}) =& \Tr[\rho(\epsilon_\text{s}) A_x(\epsilon_\text{meas}) \otimes B_y(\epsilon_\text{meas})] \\
        =& \ab(1- \frac{4\epsilon_\text{s}}{3})(1-2\epsilon_\text{meas})^2 \ab(-\hat{a}_x \cdot \hat{b}_y),
    \end{aligned}
\end{equation}
where we assume the same measurement infidelity $\epsilon_\text{meas}$ for both parties.
We then define the \emph{combined infidelity} $\epsilon$ 
\begin{equation}\label{eq:combined_infid}
    \epsilon = 1- \ab(1- \frac{4\epsilon_\text{s}}{3})(1-2\epsilon_\text{meas})^2 \simeq 4\ab(\frac{\epsilon_\text{s}}{3} + \epsilon_\text{meas}),
\end{equation}
and this leads to the quantum value with infidelity $\epsilon$ as
\begin{equation} \label{eq:omega_Q(epsilon,M)}
    \omega_\text{Q}(\epsilon, M) = \frac{1+(1-\epsilon)Q(M)}{2},
\end{equation}
where 
\begin{equation} \label{eq:Q(M)}
    Q(M) = \max_{\hat{a}_x,\hat{b}_y} \sum_{x,y}(-M_{x,y}) \hat{a}_x \cdot \hat{b}_y.
\end{equation}
In contrast, the classical value is given by 
\begin{equation} \label{eq:omega_C(M)}
    \omega_\text{C}(M) = \frac{1+C(M)}{2},
\end{equation}
with Eq.~\eqref{eq:C(M)_def}, yielding the quantum-classical gap of the nonlocal game characterized by $M$ under the noise parameter $\epsilon$ as
\begin{equation} \label{eq:LCTC_Delta_omega}
    \Delta \omega(\epsilon, M) = \frac{(1-\epsilon) Q(M) - C(M)}{2}.
\end{equation}
As an exemplary case, the CHSH game, i.e., $P(x,y) = 1/4$ and $\beta_1=\beta_2 = 0$, leads to
\begin{equation}
    M_\text{CHSH} = \frac{1}{4}\mqty{1 & 1\\ 1 & -1},
\end{equation}
resulting in $C(M_\text{CHSH}) = 1/2$ and $Q(M_\text{CHSH}) = 2\Vab{M_\text{CHSH}}_2 = 1/\sqrt{2}$, where $\Vab{M}_2$ represents the maximal singular value of $M$~\cite{Epping2013}.
Therefore, 
\begin{equation} \label{eq:Delta_omega_epsilon_CHSH}
    \Delta \omega(\epsilon, M_\text{CHSH}) = \frac{(1-\epsilon)\sqrt{2} - 1}{4},
\end{equation}
reproducing the conventional gap $(\sqrt{2}-1)/4$ in Sec.~\ref{sec:non-local_games} when $\epsilon = 0$.

With the form for $\Delta \omega$ in hand, we evaluate the gap across representative parameter regimes; we first consider zero infidelity $\epsilon = 0$ to focus on the problem structure $M$.
Fig.~\ref{fig:extended_protocols}(a) reproduces the baseline, i.e., $\beta_1 = \beta_2 = \beta,\, P(x,y)=P(x)P(y)$, and $P(x=1) = P(y=1) = p$, studied in Ref.~\cite{2407.21723}. 
Here, we mask the region $\Delta \omega < \num{e-4}$ since such a small gap will render the required rate $R_\text{req}(\alpha)$ dauntingly large, as discussed in Sec.~\ref{sec:rate_requirement}.
The quantum advantage vanishes at $\beta=0.5$; since the off‑diagonal weights $M_{01} = p(1-p)(1-2\beta_1)$ and $M_{10} = p(1-p)(1-2\beta_2)$ approach zero, an optimal strategy can be implemented based solely on its local input without requiring nonlocal correlations. 
As $\beta$ is tuned away from $0.5$, the off‑diagonal terms increase, allowing a single choice of measurement settings to support large correlators across all four input pairs, and the advantage peaks near $\beta = 0$ or 1.
Fig.~\ref{fig:extended_protocols}(b) considers correlated input distributions $P(x,y)$.
While a utility parameter $\beta$ near $0.5$ again eliminates quantum advantage, the correlation of the input distribution modifies the parameter space to exhibit large $\Delta \omega$.

\subsection{Fidelity criterion}
\label{sec:requirement_on_fidelity}

\begin{figure}[b]
    \centering
    \includegraphics[width=0.99\linewidth]{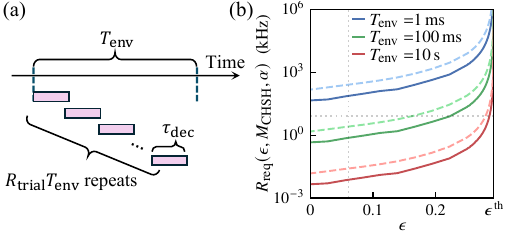}
    \caption{\textbf{Finite-statistics rate requirement under a bounded duration.}
    \textbf{(a)} In an LCTC task, each decision must be completed within the local deadline $T_{\mathrm{loc}}$, while statistically valid certification can only accumulate over a duration $T_{\mathrm{env}}$, yielding $R_{\mathrm{trial}}\,T_{\mathrm{env}}$ usable trials, where $R_\text{trial}$ represents the achievable trial rates of the LCTC tasks, usually determined by the rate of remote entanglement generation.
    \textbf{(b)} Required rate $R_\text{req}(\alpha)$ depending on the combined infidelity $\epsilon$ for the CHSH game ($\epsilon^\text{th} = 1-1/\sqrt{2} \approx 0.3$).
    The solid (dashed) line represents the significance level of $\alpha= \num{5e-2}\, (\num{e-3})$.
    The horizontal, gray dotted line represents \qty{7.9}{\kHz} and the vertical, gray dotted line represents $\epsilon = 0.061$ (see Table~\ref{table: key_parameters}).
    }
    \label{fig:finite_stats_rate}
\end{figure}

The \emph{fidelity criterion}, a threshold for the combined fidelity to achieve quantum advantage, immediately follows from the above discussion;
the expression Eq.~\eqref{eq:LCTC_Delta_omega} gives the threshold for $\epsilon$ to achieve $\Delta \omega > 0$,
\begin{equation}
\label{eq:epsilon^th}
    \epsilon^\text{th}(M) = 1- \frac{C(M)}{Q(M)} = \frac{2\Delta \omega(0,M)}{Q(M)},
\end{equation}
and the fidelity criterion is given by
\begin{align}
\label{eq:fidelity_criterion}
    \epsilon<\epsilon^\text{th}(M).
\end{align}
Fig.~\ref{fig:extended_protocols}(c) shows how the infidelity suppresses the gap in the symmetric [$P(x,y) = 1/4$ and $\beta_1=\beta_2=\beta$] and asymmetric utility case [$P(x,y) = 1/4$ and $\beta_1/2=\beta_2$].
For each $\beta_1$, the gap $\Delta \omega$ decreases monotonically with increasing $\epsilon$ and vanishes at a threshold $\epsilon^\text{th}$, above which no quantum advantage is attainable.  
The CHSH game ($\beta_1=\beta_2=0$) corresponds to $\epsilon^\text{th} = 1-1/\sqrt{2}$, whereas increasing $\beta$ pushes the game toward a smaller gap and correspondingly smaller $\epsilon^\text{th}$, meaning that only near-perfect entangled states and measurements may yield any advantage. 
We note that, while the fidelity criterion in Eq.~\eqref{eq:fidelity_criterion} mathematically assures $\Delta \omega(\epsilon, M) >0$, this does not suggest that the quantum strategy robustly exceeds any classical strategy within finite rounds, as discussed in Sec.~\ref{sec:finite_stats}.
This motivates the quantitative evaluation of the statistical certification requirement in the next section.

\subsection{Rate criterion} 
\label{sec:rate_requirement}

To achieve quantum advantage with finite rounds operating within the duration $T_\text{env}$ [see Fig.~\ref{fig:finite_stats_rate}(a)], two parties must play nonlocal games at a higher rate than $R_\text{req}(\alpha)$ in Eq.~\eqref{eq:R_dec(alpha)}.
In the generalized LCTC task with the game structure $M$ and infidelity $\epsilon$ associated with the quantum strategy, we recall [see Eqs.~\eqref{eq:omega_Q(epsilon,M)} and~\eqref{eq:omega_C(M)}]
\begin{equation}
    \omega_\text{C}(M) = \frac{1+C(M)}{2}, \quad \omega_\text{Q}(\epsilon, M) = \frac{1+(1-\epsilon) Q(M)}{2}.
\end{equation}
Since the utility function in Eq.~\eqref{eq:utility_AIC} can be interpreted as the probability that the parties are judged to win, the $p$-value is given by Eq.~\eqref{eq:pvalue_binomial}, resulting in 
\begin{equation}
    \begin{aligned}
        &p(\lceil m \omega_\text{Q}(\epsilon, M) \rceil,m) \\
        &=\sum_{k=\lceil m \omega_\text{Q}(\epsilon, M)\rceil}^{m}\binom{m}{k}[\omega_\text{C}(M)]^{\,k}[1-\omega_\text{C}(M)]^{\,m-k}.
    \end{aligned}
\end{equation}
Provided the window duration $T_\text{env}$, this gives the required rate in Eq.~\eqref{eq:R_dec(alpha)}
\begin{align}
    R_\text{req}(\epsilon,M,\alpha) =\frac{n_\text{req}(\epsilon,M,\alpha)}{T_\text{env}},
\end{align}
where 
\begin{equation} 
    n_\text{req}(\epsilon, M, \alpha) = \min\{m\in \mathbb{N}_+ \mid p(\lceil m\omega_\text{Q}(\epsilon, M)\rceil, m) < \alpha \}
\end{equation}
from Eq.~\eqref{eq:n(alpha)_def}.
Then, the \emph{rate criterion}, the criterion for the trial rate $R_\mathrm{trial}$ of the LCTC task execution to achieve the significance level $\alpha$ within the limited time window $T_\text{env}$, is
\begin{equation} \label{eq:game_play_rate_requirement}
    R_\text{trial} > R_\text{req}(\epsilon,M,\alpha) = \frac{n_\text{req}(\epsilon,M,\alpha)}{T_\text{env}},
\end{equation}
provided the fidelity criterion [Eq.~\eqref{eq:fidelity_criterion}] is satisfied.

As a quantitative benchmark, we evaluate the required rate $R_\text{req}(\epsilon, M_\text{CHSH}, \alpha)$ with several typical durations $T_\text{env}$, as shown in Fig.~\ref{fig:finite_stats_rate}(b).
This shows that the required rate increases for larger $\epsilon$, with a particularly sharp increase near the threshold $\epsilon^\text{th}$.
Such a strong dependence is expected across a wide range of game settings near the infidelity threshold, underscoring the importance of maintaining a sufficient margin above the threshold.

\subsection{Decision criterion}
\label{sec:decision_criteria}

The remaining criterion for quantum advantage in the LCTC task is that the operations required for executing the quantum strategy must complete within the finite local decision window $T_\mathrm{loc}$. 
The local decision latency $\tau_\mathrm{dec}$ consists of at least local measurement-basis selection and measurement. 
In a wide range of physical platforms for realizing a single qubit, the measurement of a $\pm 1$-valued traceless observable such as $A_x = \sum_{i} (\hat{a}_x)_i \sigma_i$ is implemented via a single-qubit rotation (gate) followed by a measurement in the standard ($Z$) basis.
Thus, the local decision latency is
\begin{equation}  \label{eq:tau_dec_decomp}
    \tau_{\mathrm{dec}} = \tau_{\mathrm{rot}} + \tau_{\mathrm{meas}}, 
\end{equation}
where $\tau_{\mathrm{rot}}$ denotes the time required for a single-qubit rotation (gate) and $\tau_{\mathrm{meas}}$ is the $Z$-measurement duration.
The decision must be made within the local decision window $T_\mathrm{loc}$; therefore, with $\tau_{\mathrm{dec}}$ in Eq.~\eqref{eq:tau_dec_decomp}, the decision criterion is given by
\begin{equation} \label{eq:lc_condition2}
    \tau_{\mathrm{dec}} < T_\text{loc}.
\end{equation}

In typical implementations of quantum network nodes, measurement duration is much longer than the single-qubit gate time; therefore, $\tau_\text{dec}$ is often dominated by $\tau_{\mathrm{meas}}$.
Still, sufficiently fast measurements are widely available across qubit platforms;
for example, the measurement of superconducting qubits can be completed in a microsecond~\cite{acharya2025quantum}, and trapped-atom qubits, such as trapped ions and neutral atoms, have a measurement time of a microsecond or less, using fast measurement techniques and cavity-assisted measurements~\cite{Falconi2025,Ma2023,Grinkemeyer2025,Deist2022PRL,Wang2025PRL}.
Photonic qubits are substantially faster, with photon pulse durations of less than a nanosecond, which can be measured in an arbitrary basis by fast optical modulation followed by photon detectors~\cite{psiquantum2025manufacturable}.
However, photonic entanglement distribution faces several challenges in meeting the three criteria simultaneously, as photon loss directly translates to coordination failure~\cite{2407.21723}.
Therefore, in the next section, we discuss the memory-based implementation of the quantum strategy for LCTC tasks and elucidate the concrete requirements to simultaneously meet all three criteria identified in this section.

\section{Hardware requirements to meet the operational criteria}
\label{sec:multiplex}

We now focus on memory-based quantum network operation for implementing quantum strategies for LCTC tasks and derive platform-independent \emph{hardware requirements} to achieve a quantum advantage based on the operational criteria identified in the previous section.
Unlike direct photon-pair distribution, memory-based operations based on heralded entanglement generation and event-ready operations can achieve a quantum-classical gap even in the presence of strong optical attenuation, thanks to the ability to share entanglement between parties before the decision-making events are triggered.
In particular, we discuss \emph{a time-multiplexed} architecture for simultaneously achieving robust event-ready operations and high trial rates, supported by the efficient use of the photonic channel.
In Sec.~\ref{sec:event_ready_protocol}, we first describe the overall memory-based event-ready protocol for executing the quantum strategy of LCTC tasks.
Then, in Sec.~\ref{sec:continuous-operation}, we discuss the time-multiplexed protocol, which is essential for satisfying the stringent rate criterion in a scalable manner.
Finally, we summarize the hardware requirements in Sec.~\ref{sec:requirement_summary} and Table~\ref{table:hardware_requirements}.

\subsection{Quantum-memory-based event-ready protocol}
\label{sec:event_ready_protocol}

\begin{figure*}[t]
    \centering
    \includegraphics[width=1\linewidth]{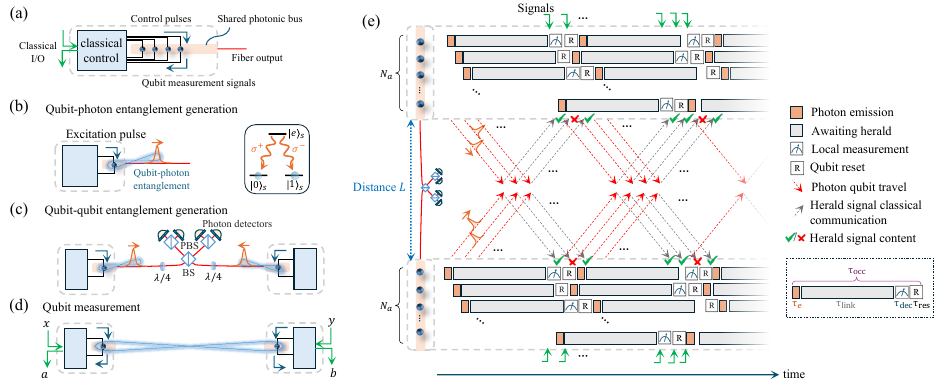} 
        \caption{\textbf{Multiplexed memory-based continuous operation protocol for LCTC tasks.}
        \textbf{(a--d)} Hardware primitives: 
        (a) a classical controller addresses a bank of memory qubits coupled to a shared photonic bus; 
        (b) an excitation pulse generates qubit-photon entanglement and emits a photon to the fiber channel; 
        (c) two photons interfere at a beamsplitter (BS), before being measured in the polarization basis by polarizing beamsplitters (PBS) and photon detectors; the detection pattern heralds the successful HEG trial;
        (d) once the entangled qubit pair is ready, upon input arrival, each node performs a local basis choice and qubit measurement to decide its action within the local decision window $T_{\mathrm{loc}}$. 
        \textbf{(e)} Pipelined time-multiplexed operation: entanglement-generation attempts are launched at intervals of $\tau_e$ across multiple memories while earlier attempts are awaiting the herald signal through classical communication over a link of length $L/2$; successful pairs are buffered and consumed on demand at coordination triggers (green arrows), followed by local measurement and reset.
        Continuous operation without stalling requires sufficient memory depth to cover the per-qubit occupancy time.}
    \label{fig:multiplexing}
\end{figure*}

We first describe the protocol for a scalable implementation of quantum strategies for LCTC tasks, a quantum-memory-based event-ready protocol, as illustrated in Fig.~\ref{fig:multiplexing}(a--d).
Central to this operation is the probabilistic heralded entanglement generation (HEG) over the photonic link~\cite{Beukers2024}. 

As a representative example of HEG, in Fig.~\ref{fig:multiplexing}(a--c), we illustrate HEG operation based on the generation of qubit-photon entanglement and two-photon interference~\cite{Beukers2024}, which is commonly adopted in experiments~\cite{Krutyanskiy2023telecom, Saha2025NC, CoveyNaturePhysics2025}.
The setup is schematically illustrated in Fig.~\ref{fig:multiplexing}(a).
Here, we first generate entanglement between a memory qubit and a photon, for example, by the direct emission of a photon into a separate \emph{mode} depending on the final qubit state, where the mode represents certain degrees of freedom of a photon, such as polarization, time bins, and frequency; in Fig.~\ref{fig:multiplexing}(b), we illustrate the case for generating a photon encoded in polarization degrees of freedom~\cite{Krutyanskiy2023telecom, Hartung2024}.
As shown in Fig.~\ref{fig:multiplexing}(c), two parties generate qubit-photon entanglement, followed by a measurement in the middle where a beamsplitter (BS) is placed for the interference of incoming photons before measuring the polarization states of the photons coming out from the output ports of the BS using polarizing beamsplitters (PBS) and photon detectors. 
The BS erases the \emph{which-path} information of the two incoming photons via two-photon interference (TPI); therefore, the photon detection projects the two remote memory qubits onto a maximally entangled state, such as $\ket{\Psi^{-}}$, depending on the measurement pattern~\cite{Simon2003}.

The overall operation cycle of the memory-based event-ready protocol is divided into the following three stages.

\begin{enumerate}
  \item \textit{Entanglement generation:} remote entanglement is established prior to the arrival of coordination inputs through repeated HEG attempts over the photonic link, with successful trials identified by a heralding signal. 
  \item \textit{Entanglement storage:}
  following a successful HEG, the Bell pair is retained in quantum memory until it is used in a coordination round. 
  The memory lifetime $\tau_\mathrm{mem}$ must be long compared to the interval between entanglement generation and its subsequent use.
  \item \textit{Local measurement and reset:} upon receiving inputs \(x\) and \(y\), the parties perform local measurements on one of their stored entangled qubits according to a pre‑agreed strategy to produce outputs $a$ and $b$ [Fig.~\ref{fig:multiplexing}(d)].  
  Because the entanglement has been prepared and stored beforehand, no inter-node communication is required after input arrival. After local measurements, the qubit is reset to be used again, starting from stage 1.
\end{enumerate}

\subsection{Pipelined time-multiplexed operation and buffering}
\label{sec:continuous-operation}

To achieve high throughput in LCTC and to satisfy the rate criterion, it is desirable for the Bell pairs to be available at a high rate.
While the sequential use of a single photonic channel with only one memory qubit leads to inefficient link utilization, as each probabilistic HEG attempt occupies the channel until the herald information is returned, time-multiplexed operation~\cite{Huie2021PRR,Li2024PRXQ,Sunami2025PRXQ} distributes successive attempts across multiple memory qubits and substantially reduces idle time, resulting in orders-of-magnitude increases in the entanglement generation rate~\cite{Huie2021PRR,Li2024PRXQ,Sunami2025PRXQ}.

The time-multiplexed protocol is illustrated in Fig.~\ref{fig:multiplexing}(e). 
We consider $N_a$ qubits that are coupled to a photonic bus connecting to a single optical channel [Fig.~\ref{fig:multiplexing}(a)].
We then perform HEG attempts sequentially, separated by the duration $\tau_e$ of each qubit-photon entanglement generation~\cite{Li2024PRXQ,Sunami2025PRXQ}.
Here, we let $\tau_\mathrm{link}$ denote the time required for the photons to reach the detection device and also the time required for the herald signals to propagate back to the memory, which is much longer than $\tau_e$ for long-distance communications typically required for LCTC, such as those illustrated in Sec.~\ref{sec:application_examples}.
In time-multiplexed operation, instead of idling the channel for $\tau_\mathrm{link}$ in the case of $N_a=1$, the channel usage is spread over $N_a\gg1$ HEG trials [Fig.~\ref{fig:multiplexing}(e)].
After receiving the herald signal, which will arrive with an interval of $\tau_e$ for $N_a$ qubits, the successfully entangled qubits are stored while others are reset for immediately performing another round of HEGs. \@
Simultaneously, as the local signal $(x,y)$ arrives, the stored qubits are measured to enable quantum strategies, followed by qubit reset and the next HEG trials.

This protocol allows for continuous and highly efficient operations, with an upper bound on the HEG trial rate of $1/\tau_e$.
To see how the upper bound is saturated, consider that the single memory qubit is occupied between the launches of HEG trials; assuming that the input arrives as the entanglement becomes available for measurement, the occupancy time is given by
\begin{equation}  \label{eq:occ_time}
    \tau_\text{occ} = \tau_e+\tau_\text{link} + \tau_\text{dec} + \tau_\text{res},
\end{equation}
where $\tau_\text{res}$ represents the qubit reset time.
The launch of HEG trials at intervals of $\tau_e$ revisits the same qubit no sooner than $N_a \tau_e$.
Thus, a number of qubits $N_a$ satisfying 
\begin{equation} \label{eq:continuous_operation}
    N_a \tau_e \geq \tau_\text{occ}
\end{equation}
saturates the upper bound of the HEG trial; i.e., channel idle time is eliminated by time multiplexing.
In general, the HEG attempt rate is given by
\begin{equation}
\label{eq:Gamma_HEG}
    \Gamma_\text{HEG} =  \min\ab(\frac{1}{\tau_e}, \frac{N_a}{\tau_\text{occ}}).
\end{equation}
Furthermore, the coherence time of the memory qubit $\tau_\mathrm{mem}$ must be sufficiently longer than the occupancy time.
To be more concrete, we model that the error accumulation during the occupancy time $\tau_\text{occ}$ increases the entanglement infidelity to 
\begin{equation}
     \epsilon_\text{s}' = \epsilon_\text{s} + 2(1-e^{-\tau_\text{occ}/\tau_\text{mem}}),
\end{equation}
where the factor of 2 comes from the fact that memory decoherence affects both qubits of the Bell pair.
Therefore, we require that the memory-induced errors do not increase the combined infidelity $\epsilon$ [Eq.~\eqref{eq:combined_infid}] above $\epsilon^\mathrm{th}$, i.e.,~
\begin{equation}\label{eq:memory_requirement}
    \tau_\text{mem} > \tau_\text{mem}^\text{th}(\tau_\text{occ}, \epsilon^\text{th}, \epsilon_\text{meas}, \epsilon_\text{s}),
\end{equation}
where
\begin{equation}
    \begin{aligned}
        &\frac{\tau_\text{mem}^\text{th}(\tau_\text{occ}, \epsilon^\text{th}, \epsilon_\text{meas}, \epsilon_\text{s})}{\tau_\text{occ}} \\
        &= - \ab\{\ln\ab(1-\frac{3}{8}\ab[1-\frac{4\epsilon_\text{s}}{3} -\frac{1-\epsilon^\text{th}}{(1-2\epsilon_\text{meas})^2}] ) \}^{-1}.
    \end{aligned}
\end{equation}

Finally, to see how the rate criterion is satisfied, consider the success probability of the HEG trials, $p_\text{ent}(L)$, limited by the inherently probabilistic nature of the HEG protocol, channel losses over distance $L$, and other factors including limited detector efficiency. 
Then, the total rate of successfully generating entanglement, which we call an \emph{HEG rate}, is
\begin{equation} \label{eq:R_EG}
    R_\text{HEG}(L) = N_\text{ch} p_\text{ent}(L)\Gamma_\text{HEG},
\end{equation}
where we added $N_\text{ch}$ as the multiplicative factor for channel multiplexing, i.e., the parallel use of $N_\text{ch}$ networking hardware and optical channels.
In the memory-based protocol discussed here, $R_\text{HEG}(L)$ directly translates to the trial rate $R_\text{trial}$ of Eq.~\eqref{eq:game_play_rate_requirement}.
To meet the rate criterion of Eq.~\eqref{eq:game_play_rate_requirement}, with the combined infidelity $\epsilon$ [which we assume satisfies the fidelity criterion, Eq.~\eqref{eq:fidelity_criterion}], the HEG rate must satisfy
\begin{equation} \label{eq:R_EG_requirement}
    R_\text{HEG}(L) > R_\text{req}(\epsilon,M,\alpha).
\end{equation}

\subsection{Hardware requirements for time-multiplexed, memory-based protocol}
\label{sec:requirement_summary}

\begin{table}   
    \caption{Important parameters of LCTC tasks and memory-based event-ready protocols for implementing quantum strategies, and the requirements to meet operational criteria for achieving certified quantum advantage. 
    }
    \small
    \begin{ruledtabular}
        \begin{tabular}{lccc}
            \multicolumn{4}{c}{Key parameters of LCTC tasks and hardware} \vspace{3pt} \\
             & Parameter & Symbol & Ref. \\
            \hline
            \multirow{5}{*}{\makecell[l]{LCTC}} & Game structure & $M$ & Eq.~\eqref{eq:M_XOR} \\
            & Decision window & $T_\text{loc}$ & Sec.~\ref{sec:lc_tacit_coord}\\
            & Communication time & $T_\text{comm}$ & Sec.~\ref{sec:lc_tacit_coord} \\
            & Stationary window & $T_\text{env}$ & Sec.~\ref{sec:lc_tacit_coord} \\
            & Target $p$-value & $\alpha$ & Sec.~\ref{sec:finite_stats} \\
            \hline
            \multirow{6}{*}{\makecell[l]{Hardware}}
            & Entanglement infidelity& $\epsilon_\text{s}$ & Eq.~\eqref{eq:werner} \\
            & Measurement infidelity & $\epsilon_\mathrm{meas}$ & Eq.~\eqref{eq:A_x(epsilon)} \\
            & Combined infidelity & $\epsilon$ & Eq.~\eqref{eq:combined_infid} \\
            & HEG rate & $R_\mathrm{HEG}$ & Eq.~\eqref{eq:R_EG} \\
            & Decision time & $\tau_\mathrm{dec}$ & Eq.~\eqref{eq:tau_dec_decomp} \\
            & Memory lifetime & $\tau_\mathrm{mem}$ & Sec.~\ref{sec:continuous-operation} \\
        \end{tabular}
        \vspace{10pt}
        \begin{tabular}{lcc}
            \multicolumn{3}{c}{Operational criteria and hardware requirements} \vspace{3pt} \\
            Criteria & hardware requirement & Ref. \\
            \hline
            Fidelity criterion & $1-\epsilon > \dfrac{C(M)}{Q(M)}$ & Eq.~\eqref{eq:fidelity_criterion} \\
            Rate criterion & $R_\text{HEG} > \dfrac{n_\text{req}(\epsilon, M, \alpha)}{T_\text{env}}$ & Eq.~\eqref{eq:game_play_rate_requirement} \\
            Decision criterion & $\tau_\text{dec} < T_\text{loc} $ & Eq.~\eqref{eq:lc_condition2} \\
        \end{tabular}
    \end{ruledtabular}
    \label{table:hardware_requirements}
\end{table}

The above discussions identify requirements for the hardware implementation of the LCTC tasks, which we summarize below and in Table~\ref{table:hardware_requirements}.
First, the fidelity criterion [Eq.~\eqref{eq:fidelity_criterion}] must be satisfied by high-fidelity remote entanglement, long coherence time [Eq.~\eqref{eq:memory_requirement}], and high-fidelity measurement.
Second, the rate criterion [Eq.~\eqref{eq:game_play_rate_requirement}] requires sufficiently high-rate HEG, which can be achieved by several factors, mainly short $\tau_e$, high $p_\mathrm{ent}$, and large $N_a$ (see Sec.~\ref{sec:continuous-operation}).
Third, the decision criterion [Eq.~\eqref{eq:lc_condition2}] is satisfied by fast qubit measurement with measurement-basis rotation.

\section{Time-multiplexed quantum network node with neutral atoms and optical cavities}\label{sec:resource_estimation}

Atomic qubits, such as trapped ions~\cite{Brown2016} and neutral atoms trapped in optical tweezer arrays~\cite{Kaufman2021}, constitute scalable and high-performance platforms that support both large-scale quantum computing~\cite{Bluvstein2024,helios2025} and memory-based quantum network operations~\cite{vanLeent2022Nature,Liu2026}.
When combined with optical cavities that substantially enhance atom-photon coupling, these systems enable high-rate, high-fidelity quantum networking~\cite{Main2025Nature, Saha2025NC,Li2024PRXQ, Sunami2025PRXQ, 2507.01229}, fast qubit measurement~\cite{Deist2022PRL,Wang2025PRL}, and reset~\cite{Wang2025}.
In light of these prospects, in this section, we propose a concrete hardware implementation that meets the stringent operational criteria and resulting hardware requirements identified above (summarized in Table~\ref{table:hardware_requirements}), which is expected to support quantum advantage in a wide variety of LCTC tasks.
First, in Sec.~\ref{sec:telecom_node}, we introduce the overall system:  $^{171}\text{Yb}$ atoms coupled to an optical cavity.
In Sec.~\ref{subsec:execute_LCTC}, we present protocols for cavity-enhanced remote entanglement generation and fast qubit measurement, along with detailed numerical modeling of the performance.
Based on these results, we then project the overall performance for realistic metropolitan-scale coordination and compare it with the operational criteria in Sec.~\ref{subsec:overall_performance}.
Finally, in Sec.~\ref{subsec:prospect_other_hardware}, we discuss the prospects for further enhancement, as well as other possible hardware implementations.

\subsection{Time-multiplexed, telecom-band quantum network node}
\label{sec:telecom_node}

\begin{figure}[t]
    \centering
    \includegraphics[width=0.9\linewidth]{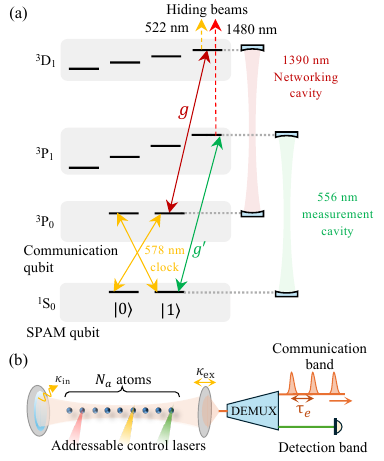}
    \caption{\textbf{Telecom-band multiplexed network node with ${}^{171}\mathrm{Yb}$ atoms and optical cavities.}
    \textbf{(a)} Level diagram of ${}^{171}\mathrm{Yb}$ atoms;
    the state preparation and measurement (SPAM) qubit is encoded in the ${}^{1}\text{S}_0$ manifold, with the metastable ${}^{3}\text{P}_0$ state serving as the communication qubit.
    A telecom ``networking'' cavity resonant at \qty{1390}{\nm} provides the atom-photon interface for remote entanglement, while a separate ``measurement'' cavity resonant at 556\,nm enables fast cavity-assisted readout and reset; auxiliary laser beams (e.g., 578\,nm clock light and 522, 1480 nm hiding beams) support coherent control and site-selective protection of spectator atoms.
    \textbf{(b)} Time-multiplexed multi-atom operation: addressable control pulses sequentially couple $N_a$ atoms to generate a train of communication-band photons at intervals $\tau_e$; wavelength demultiplexing routes telecom photons to the network (Communication band) and measurement photons to local detection (Detection band).
    }
    \label{fig:yb-impl}
\end{figure}

    Among a wide variety of atomic species used for quantum information processing and quantum networking, neutral ${}^{171}$Yb atoms are particularly well suited for LCTC tasks, owing to long coherence times~\cite{Jenkins2022PRX}, high-fidelity qubit control~\cite{Jenkins2022PRX, Ma2022, Ma2023}, access to metastable-state qubit manifolds~\cite{Lis2023}, and compatibility with high-rate, high-fidelity remote entanglement generation at the telecom band~\cite{Huie2021PRR,Li2024PRXQ,Sunami2025PRXQ}.
    In particular, we leverage the availability of ground-state and metastable-state qubits to enable parallel wavelength-multiplexed operation. 
   
    As illustrated in Fig.~\ref{fig:yb-impl}(a), the node employs two optical cavities resonant at \qty{556}{\nm} and \qty{1390}{\nm} dedicated to state preparation and measurement (SPAM) and telecom-band quantum networking, respectively. The \qty{556}{\nm} cavity addresses the ${}^{1}\text{S}_{0}\leftrightarrow{}^{3}\text{P}_{1}$ cycling transition to enable fast, high-fidelity fluorescence readout, while the \qty{1390}{\nm} cavity couples to the ${}^{3}\text{P}_{0}\leftrightarrow{}^{3}\text{D}_{1}$ transition to generate telecom-band photons for remote entanglement. Dual-wavelength cavity operation can be realized using crossed-cavity geometries~\cite{Brekenfeld2020}, nanophotonic cavities incorporating multiple narrow-band in-fiber Bragg-grating mirror pairs~\cite{Sunami2025PRXQ}, or free-space cavities resonant at both \qty{556}{\nm} and \qty{1390}{\nm}, enabled by multilayer dielectric coatings, facilitated by the nearly $2.5\times$ wavelength separation between the two transitions~\cite{Garcia2018}. Coherent mapping between the ground-state and metastable-state qubits is performed via the narrow-line clock transition ${}^{1}\text{S}_{0}\leftrightarrow{}^{3}\text{P}_{0}$, which enables rapid and high-fidelity coherent transfer of quantum states between the two qubit manifolds with fidelities at the $99.9\%$ level on microsecond timescales~\cite{Lis2023}.
    
    A crucial requirement is that the cavity hosts a large number of atoms to enable the time-multiplexed operations discussed in
    Sec.~\ref{sec:continuous-operation}. This requirement can be met across several cavity platforms: free-space cavities, such as Fabry-Perot cavities~\cite{aqua2025} and bowtie cavities~\cite{Li2024PRXQ}, feature large mode volumes that can couple hundreds of atoms, while a nanofiber cavity with a long cavity mode also allows a comparable number of atoms to be coupled~\cite{Sunami2025PRXQ}. In such many-atom cavity systems, it is essential to suppress errors arising from atom-atom crosstalk mediated by strong coupling to a shared cavity mode. This error suppression can be achieved using site-resolved \textit{hiding} laser beams that are near-resonant to the transitions originating from the relevant excited states of the cavity-coupled transitions~\cite{Hu2025PRL,Li2024PRXQ,2507.01229}, for example, at wavelengths near \qty{1480}{\nm} and \qty{522}{\nm} corresponding to the ${}^{3}\text{P}_{1}$ and ${}^{3}\text{D}_{1}$ manifolds, respectively. When focused on individual atoms and applied selectively by using acousto-optic deflectors or integrated photonic switches, these beams shift the transitions of selected atoms out of cavity resonance, thereby allowing selective atom-cavity coupling with negligible time cost of switching~\cite{Hu2025PRL, Zhao2025, Wei2026}.

\begin{figure*}[t]
    \centering
    \includegraphics[width=0.99\linewidth]{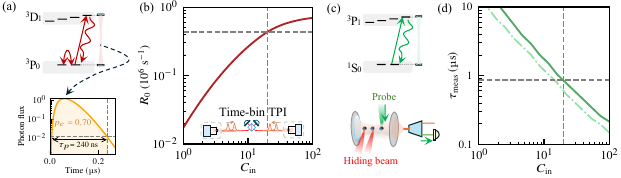}
    \caption{\textbf{Performance of the telecom-band Yb atom-cavity network nodes.}
    \textbf{(a)} Telecom-band atom-photon entanglement generation.
    Initially, the atom is in an equal superposition of two sublevels of ${}^{3}\mathrm{P}_0$ manifold (communication qubit).
    The atom in $\ket{^{3}\text{P}_0,m_F = 1/2}$ is then excited by fast laser pulse (straight arrow), followed by cavity-enhanced photon emission that subsequently brings the atomic state to the ${}^{3}\mathrm{P}_0$ manifold (wiggly arrow).
    Repeating this procedure after the state swap ($\pi$ pulse) on the communication qubit, a maximally entangled atom-photon qubit pair is generated where the photonic qubit is encoded in the time-bin basis.
    Also shown is the numerically simulated photon flux, which leaks out to the fiber channel for remote entanglement generation. 
    The pulse width $\tau_p$ is the time at which the flux decays to \num{e-2} of its maximum value, when the next entanglement generation trial with another atom can be initiated.
    A simulation result for $C_\text{in}=20$ is shown, for which $\tau_p = \qty{240}{\ns}$ and $p_e = 0.70$, resulting in TPI success probability of $p_e^2/2=0.25$.
    \textbf{(b)} Intrinsic HEG rate (see text) as a function of $C_\text{in}$ at TPI infidelity 
    below 1\% (see Appendix~\ref{app:Bell_state_TPI} for the details).
    The vertical dashed line is at $C_\text{in}=20$, where $R_0 = \qty{4.3e5}{\s^{-1}}$.
    \textbf{(c)} Cavity-enhanced qubit measurement. 
    Population in $\ket{^{1}\text{S}_0, m_F=1/2}$ is weakly excited to $\ket{^{3}\text{P}_1, m_F=3/2}$ by excitation laser, followed by photon emission into the cavity mode that is then directed to the detector for state discrimination.
    {\textbf{(d)}} Performance of cavity-assisted fluorescence readout.
    We evaluate the measurement time $\tau_\text{meas}$ that achieves the measurement infidelity of $\epsilon_\text{meas} = 0.002$ (solid) and $\epsilon_\text{meas} = 0.01$ (dash-dotted) (see Appendix~\ref{app:cavity_readout} for the details).
    For $C_\text{in}=20$ (vertical dashed line), 0.2\% measurement error is obtained by $\tau_\text{meas} = \qty{870}{\ns}$, leading to $\tau_\text{dec} = \qty{1}{\us}$ in Eq.~\eqref{eq:tau_dec_decomp} with the qubit-rotation time $\tau_\text{rot} = \qty{100}{\ns}$.
    }
    \label{fig:resource_estimation}
\end{figure*}

\subsection{Executing LCTC tasks with a cavity-coupled neutral atom array} \label{subsec:execute_LCTC}

In the following, we provide a detailed analysis of cavity-assisted remote entanglement generation and qubit measurement (Fig.~\ref{fig:resource_estimation}), which determines the critical performance parameters achievable with the proposed hardware.
In Sec.~\ref{sec:V_A1}, we first present an example implementation of HEG with $^{171}\text{Yb}$ atoms. 
In Sec.~\ref{sec:V_A2}, we evaluate a cavity-enhanced qubit measurement, showing that the measurement is sufficiently fast to meet the decision criterion [Eq.~\eqref{eq:lc_condition2}].
        
\subsubsection{Remote entanglement generation} \label{sec:V_A1}

Here, we analyze the performance of remote entanglement generation assisted by the networking cavity, as summarized in Fig.~\ref{fig:resource_estimation}(a). Here, we employ time-bin encoding for telecom-band photon~\cite{CoveyNaturePhysics2025,Kikura2025_taming, Saha2025NC}, which is well suited for long-distance fiber communication owing to its robustness against polarization and phase fluctuations~\cite{Huie2021PRR, Knaut2024Nature}.
In each attempt, an atom-photon entangled state is prepared by (i) initializing the atom in a coherent superposition between, e.g., the two communication qubit states, (ii) applying a fast excitation pulse to generate an ``early'' photon with a pulse width $\tau_p$, (iii) performing a coherent internal-state population swap, and (iv) applying a second excitation pulse to generate a ``late'' photon.  
Then, remote atom-atom entanglement is achieved by interfering and measuring the photons emitted from two distant nodes at a midpoint~\cite{Barrett2005,Sunami2025PRXQ,Huie2021PRR}. 
This protocol implements the TPI scheme described in Sec.~\ref{sec:event_ready_protocol}, where subsequent detection events herald the generation of remote atom-atom Bell pairs~\cite{Barrett2005}. 

The atom-photon entanglement generation probability $p_e$ determines the upper bound of the remote atom-atom entanglement generation success probability, which is given by $p_e^2/2$; 
the factor $1/2$ comes from the upper bound of the success probability of entanglement swapping via TPI~\cite{Calsamiglia2001}.
Since two photon pulses in separate time bins are involved in a single trial, the intrinsic HEG rate is
\begin{equation}
\label{eq:R_0}
    R_0 = \frac{p_{e}^2}{2} \frac{1}{\tau_e},
\end{equation}
where the HEG trial period
\begin{align}
\label{eq:tau_e}
    \tau_e = 2\tau_p + \tau_\text{swap},
\end{align}
where $\tau_\text{swap}$ represents the qubit state swap ($\pi$ pulse) time between the two time bins.
In our atom-cavity emitter, the performance can be quantified by a single parameter, the internal cooperativity $C_{\mathrm{in}}=g^{2}/(2\kappa_\text{in}\gamma)$~\cite{Goto2019PRA}, where $g$ is the atom-cavity coupling rate, $\gamma$ is the spontaneous decay rate from the excited state, and $\kappa_{\mathrm{in}}$ denotes the intrinsic cavity decay rate. 
We numerically simulate the atom-cavity system dynamics during photon generation, as shown in Fig.~\ref{fig:resource_estimation}(a);
this gives both the emission probability $p_e$ and the pulse width $\tau_p$ for a given atom-cavity system parameter, thereby providing the intrinsic HEG rate $R_0$.
Fig.~\ref{fig:resource_estimation}(b) shows $R_0$ as a function of $C_\mathrm{in}$.
A larger $C_{\mathrm{in}}$ increases the probability of photon emission $p_e$ and decreases the pulse width $\tau_p$, leading to a substantially larger $R_0$.

\subsubsection{Fast qubit measurement and reset}
\label{sec:V_A2} 
We next characterize fast qubit measurement enabled by the \qty{556}{\nm} cavity coupled to the cycling transition ${}^{1}\text{S}_{0}\leftrightarrow{}^{3}\text{P}_{1}$ [see Fig.~\ref{fig:resource_estimation}(c)], which dominantly sets the local decision latency $\tau_\mathrm{dec}$. 
The primary figure of merit is the tradeoff between measurement fidelity and probe duration.
Since the state discrimination is performed by collecting the photon and distinguishing whether the fluorescence is present, longer collection times typically improve the measurement fidelity.
The cavity supports fast measurement by enhancing the photon emission rate and also by guiding the emitted photon to a well-defined mode, where the time-fidelity tradeoff can be improved by higher-quality cavities with less photon loss;
Fig.~\ref{fig:resource_estimation}(d) shows the measurement times required to reach 0.2\% (solid line) and 1\% (dash-dotted line) infidelity as a function of the cavity quality $C_\mathrm{in}$.
The model captures the competition between rapid photon scattering from the bright state and rare error processes such as dark counts and depumping. 
For each probe duration, the photon-count threshold is optimized to minimize the total misclassification probability, yielding a well-defined fidelity-time curve (see Appendix~\ref{app:cavity_readout} for details). 
As a result, even for moderate $C_\mathrm{in}$, sub-microsecond measurement times are sufficient to reach below 1\% measurement infidelity.

Finally, qubit reset for the next attempt contributes an additional time $\tau_{\mathrm{res}}$ to the per-qubit occupancy time in Eq.~\eqref{eq:occ_time}. Optical pumping-based reset of the nuclear-spin state can be performed on the order of microseconds, remaining small compared to, e.g., the communication latency at metropolitan distances. 
This reset time can be further reduced using cavity-assisted optical pumping, which enhances the scattering rate and shortens the reinitialization cycle~\cite{Wang2025}. 

\begin{table*}[htbp]
\caption{Key parameters and expected performance for continuous operations of time-multiplexed neutral-atom network nodes at metropolitan-scale distance. 
\label{table: key_parameters}}
\centering
\small
\begin{ruledtabular}
    \begin{tabular}{@{}lllll@{}}
     & Quantity & Symbol & Value & Notes \\
    \midrule
    \multirow{5}{*}{\makecell[l]{Atom-cavity\\system}}
    & Internal cooperativity & $C_{\mathrm{in}}$ & $20$ & Near-term single-atom internal cooperativity~\cite{Goto2019PRA} \\
    & Atom number in cavity mode & $N_a$ & 250 & Trapped atoms coupled to an optical cavity~\cite{Sunami2025PRXQ,Li2024PRXQ} \\
    & Qubit-rotation time & $\tau_\text{rot}, \tau_\text{swap}$ & \qty{100}{\ns} & Fast Raman gates~\cite{Lis2023, Jenkins2022PRX} for single-qubit rotation  \\
    & Qubit reset time 
    & $\tau_\text{res}$ 
    & \qty{1}{\us}  %
    & Optical repumping time~\cite{Li2024PRXQ} \\
    & Coherence time & $ \tau_\text{mem} $ & \qty{7.9}{\s} & $T_2$ relaxation time~\cite{Jenkins2022PRX}\\
    \midrule
    \multirow{4}{*}{\makecell[l]{Entanglement\\generation}}
    & Photon emission probability & $p_e$ & 0.70 & Fig.~\ref{fig:resource_estimation} \\
    & TPI success probability & $p_e^2/2$ & 0.25 & Fig.~\ref{fig:resource_estimation} \\ 
    & Photon pulse duration & $\tau_{p}$ & \qty{240}{\ns} 
    & Fig.~\ref{fig:resource_estimation} \\
    & HEG rate upper bound & $R_0$ & \qty{4.3e5}{\s^{-1}} & $p_e^2/2\tau_e$ in Eqs.~\eqref{eq:R_0} and~\eqref{eq:tau_e} \\
    \midrule
    \multirow{5}{*}{\makecell[l]{Network}}
    & Distance & $L$ & \qty{50}{\km} & Metropolitan-scale separations \\
    & Link transmission& $\eta_{\mathrm{att}}$  & 0.06 & $10^{-\alpha_\text{att} L/10}$, with $\alpha_\text{att}=$ 0.25 dB/km~\cite{FujikuraLWP}  \\
    & Detector efficiency & $\eta_{\mathrm{det}}$ & $0.9$ & Superconducting nanowire single-photon detectors~\cite{Venza2025apl}\\
    & Lumped optics efficiency & $\eta_{\mathrm{misc}}$ & $0.8$ & Couplers, filters, connectors; representative~\cite{Huie2021PRR}\\
    & Link latency 
    & $\tau_{\mathrm{link}}$ 
    & \qty{240}{\us}
    & propagation time $L/v_g$, where $v_g=$\qty[per-mode=symbol]{2.1e8}{\m\per\s} \\
    \midrule
    \multirow{8}{*}{\makecell[l]{Performance}}
    & Single-channel HEG rate &  $R_{\mathrm{HEG}}$ & \qty{7.9e3}{\s^{-1}} & $p_\text{ent}(L)\Gamma_\text{HEG}$ in Eq.~\eqref{eq:R_EG_TPI}  \\
    & HEG trial period 
    & $\tau_{e}$ 
    & \qty{580}{\ns} 
    & $2\tau_p + \tau_\text{swap}$ [Eq.~\eqref{eq:tau_e}] \\
    & $Z$-measurement time & $\tau_\text{meas}$ & \qty{870}{\ns} & Cavity-enhanced measurement (Fig.~\ref{fig:resource_estimation}) \\
    & Local decision latency 
    & $\tau_{\mathrm{dec}}$ 
    & \qty{1}{\us} 
    & $\tau_\text{meas} + \tau_\text{rot}$ in Eq.~\eqref{eq:tau_dec_decomp} (Fig.~\ref{fig:resource_estimation}) \\
    & Per-qubit occupancy time %
    & $\tau_{\mathrm{occ}}$ 
    & \qty{244}{\us}
    & $ \tau_e +\tau_{\mathrm{link}}+\tau_{\mathrm{dec}}+\tau_\text{res}$ in Eq.~\eqref{eq:occ_time} \\
    & Entanglement infidelity & $\epsilon_\text{s}$ & $<4
    $\% & Fig.~\ref{fig:resource_estimation}, channel errors~\cite{Bersin2024}, and dark count effects [Eq.~\eqref{eq:epsilon_s_bound}] \\ 
    & Measurement infidelity & $\epsilon_\mathrm{meas}$ & 0.2\% & Fig.~\ref{fig:resource_estimation} \\ 
    & Combined infidelity & $\epsilon$ & $<6.1$\% & Eq.~\eqref{eq:epsilon_bound} \\
    \midrule
    \multirow{3}{*}{\makecell[l]{Operational \\criteria}}
    & Decision criterion &  & \checkmark  & Eq.~\eqref{eq:lc_condition2}; for local decision window $T_\mathrm{loc} > \qty{1}{\us}$ \\
    & Fidelity criterion &  & \checkmark & Eq.~\eqref{eq:fidelity_criterion}; for $\epsilon^\mathrm{th} > 6.1$\% [Fig.~\ref{fig:extended_protocols}(c)] \\
    & Rate criterion &  & \checkmark & Eq.~\eqref{eq:R_EG_requirement}; for $T_\mathrm{env} > \qty{10}{\ms}$ (Fig.~\ref{fig:finite_stats_rate})\\
    \end{tabular}
\end{ruledtabular}
\end{table*}

\subsection{Overall performance and compatibility with applications to LCTC tasks} \label{subsec:overall_performance}

To evaluate the system-level performance, we now incorporate additional factors that determine the overall HEG rates, referring to the discussion in Sec.~\ref{sec:multiplex}.
First, the total HEG success probability at a distance $L$, incorporating various optical losses, is
\begin{equation}
\label{eq:p_ent}
    p_\text{ent}(L) = \frac{p_e^2}{2}\eta_\text{att}(L) \eta_\text{det}^2 \eta_\text{misc}^2,
\end{equation}
where $\eta_\text{att}(L) = 10^{-\alpha_\text{att} L/10}$ accounts for fiber attenuation with the attenuation coefficient $\alpha_\text{att}$, $\eta_{\mathrm{det}}$ is the detector efficiency, and $\eta_{\mathrm{misc}}$ accounts for additional transmission losses such as intercomponent coupling and filtering losses~\cite{Huie2021PRR}.
Therefore, substituting Eqs.~\eqref{eq:Gamma_HEG},~\eqref{eq:R_0}, and~\eqref{eq:p_ent} into Eq.~\eqref{eq:R_EG}, we obtain the total HEG rate
\begin{equation} \label{eq:R_EG_TPI}
    R_\text{HEG}(L) = N_\text{ch}R_0 \eta_\text{att}(L) \eta_\text{det}^2 \eta_\text{misc}^2 \min\ab(1, \frac{N_a \tau_e}{\tau_\text{occ}}).
\end{equation}

As a representative distance scale, we consider a $L=\qty{50}{\km}$ fiber link, similar to the cases discussed in Sec.~\ref{sec:application_examples}. 
We further assume that each node has atom-cavity systems characterized by a moderate system quality of $C_{\mathrm{in}} = 20$ for both communication and measurement cavities.
As shown in Fig.~\ref{fig:resource_estimation}(d), $C_{\mathrm{in}} = 20$ is already sufficient for a measurement duration of a microsecond that satisfies the decision criterion in Eq.~\eqref{eq:lc_condition2} for a local decision window of microseconds to \qty{10}{\us}, such as those discussed in Sec.~\ref{sec:application_examples}.

For the Bell pair infidelity $\epsilon_\text{s}$, our analysis in Fig.~\ref{fig:resource_estimation}(b) shows that the intrinsic infidelity arising from the emitted photon's impurity~\cite{Krutyanskiy2023} can be kept below the 1\% level.
For long-distance communication over $L=\qty{50}{\km}$, other factors such as phase fluctuations, temporal mode mismatch, and dark counts may have a major impact, which we briefly analyze here while leaving a more detailed modeling for future exploration.
For this analysis, Ref.~\cite{Bersin2024} provides a comprehensive benchmark of the effect of a 50-km fiber network on time-bin photons: the reported noises are optical frequency fluctuations of several kHz, polarization drift of less than a radian per second, and arrival timing precision on the order of nanoseconds.
While the effect of reported frequency noise and polarization drifts is expected to be small for our time-bin-based protocol, the arrival timing precision is expected to induce a temporal mismatch error of up to a few percent.
We then assume $D=$~\qty{10}{\Hz} as a representative dark count rate of widely adopted superconducting-nanowire single-photon detectors, while demonstrations of rates orders of magnitude better exist~\cite{Venza2025apl}. The upper bound $p_\mathrm{false}$ of the probability of false positives is estimated by dividing the dark count probability $4\tau_p D$ by the HEG success probability $p_\mathrm{ent}(L)$, i.e.,
\begin{align}
\label{eq:p_faulse}
    p_\mathrm{false}=\frac{4\tau_p D}{p_\mathrm{ent}(L)},
\end{align}
where the factor $4$ appears because two detectors accept the detection signals during the time window of $\tau_p$ for two time bins.
According to Fig.~\ref{fig:resource_estimation}, we take $\tau_p= \qty{240}{\ns}$; then, the false positive probability $4\tau_p D= 4\times(\text{240 ns})\times \text{(10 s}^{-1}) = 9.6 \times 10^{-6}$. 
With the HEG success probability
\begin{align}
\label{eq:p_ent_L}
    p_\mathrm{ent}(L=\qty{50}{\km}) = 7.7\times10^{-3},
\end{align}
from $p_e=0.7$, $\eta_{\mathrm{att}}(L = \qty{50}{\km}) = 0.06$ and assuming $\eta_{\mathrm{det}} = 0.9$ and $\eta_{\mathrm{misc}} = 0.8$ (see Table~\ref{table: key_parameters}).
Due to Eq.~\eqref{eq:p_faulse}, this gives $p_\mathrm{false}=0.12$\% 
from dark counts.
Overall, we expect a few percent errors in the remote Bell pairs generated by the near-term implementation of the time-bin TPI protocol over a 50-km fiber network with a detector placed at the midpoint.
Therefore, with an additional margin, we set the entanglement infidelity
\begin{align}
\label{eq:epsilon_s_bound}
    \epsilon_\text{s} < 0.04.
\end{align}
For the measurement error $\epsilon_\mathrm{meas} = 0.002$ in Fig.~\ref{fig:resource_estimation}, the combined error [Eq.~\eqref{eq:combined_infid}] is
\begin{align}
\label{eq:epsilon_bound}
    \epsilon < 0.061.
\end{align}
With this $\epsilon$, the estimated performance satisfies the fidelity criterion for a wide variety of XOR game instances, as shown in Fig.~\ref{fig:extended_protocols}(c).

At $C_{\mathrm{in}} = 20$ and $L= \qty{50}{\km}$, we have $\tau_e = 2\tau_p + \tau_\text{swap} = \qty{580}{\ns}$, $\tau_\mathrm{dec}=\qty{1}{\us}$, and $\tau_\text{link}= L/v_g = \qty{240}{\us}$ (Table~\ref{table: key_parameters}). 
Furthermore, considering $\tau_\mathrm{res}=\qty{1}{\us}$, the per-qubit occupancy time in Eq.~\eqref{eq:occ_time} is $\tau_\text{occ} = \qty{244}{\us}$, dominated by $\tau_\text{link}$. 
Therefore, $N_a \tau_e/\tau_\text{occ} = 0.59 <1$ for $N_a = 250$~\cite{Sunami2025PRXQ,Li2024PRXQ}. 
For this $\tau_\text{occ}$, the coherence time of the ${}^{171}$Yb qubits, on the order of seconds~\cite{Jenkins2022PRX}, is orders of magnitude longer than the required memory coherence time [Eq.~\eqref{eq:memory_requirement}], making the decoherence effect negligible.
The total HEG rate in Eq.~\eqref{eq:R_EG_TPI} is $R_\text{HEG}(L = \qty{50}{\km}) = N_a p_\text{ent}(L)/\tau_\text{occ} = \qty{7.9e3}{\s^{-1}}$ without channel multiplexing, i.e., $N_\text{ch} = 1$.
This rate surpasses the one required to achieve quantum advantage with a significance level of $\alpha=0.05$ for $T_\text{env} = \qty{100}{\ms}$ by a sufficient margin [see Fig.~\ref{fig:finite_stats_rate}(b); for the above $R_{\mathrm{HEG}}$, the rate criterion is satisfied up to $\epsilon \approx 0.2$], as well as for $T_\text{env} = \qty{10}{\ms}$ (for $\epsilon < 0.08$), supporting a robust quantum advantage for the application examples discussed in Sec.~\ref{sec:application_examples}.

\subsection{Prospect for further scaling and comment on other hardware implementations} \label{subsec:prospect_other_hardware}

The performance of the proposed network node can be enhanced in various ways, enabling both higher entanglement rates and longer communication distance.
First, while we focused on the widely adopted TPI protocol for HEG, the cavity-assisted photon-scattering (CAPS) protocol~\cite{2507.01229,Duan2004} can be used instead, which features a higher success probability, high fidelity, and robustness to a wider range of imperfections and parameter fluctuations. 
In particular, because the CAPS-based protocol for remote entanglement generation does not rely on the interference of photons coming from the two distant nodes, the infidelity arising from the arrival timing jitter can be eliminated.
A detailed analysis of the time-multiplexed CAPS operation for HEG is provided in Refs.~\cite{2507.01229, Ji2026}.
Furthermore, the achievable memory depth $N_a$ can be substantially improved by incorporating the continuous operation techniques recently demonstrated for neutral-atom quantum processors~\cite{chiu2025, Li2025}, whereby a constant supply of atoms can be interfaced with the photonic channel at a dedicated \emph{network zone}~\cite{Sunami2025PRXQ, Sinclair2025PRR}. 
Instead of hosting the qubits in the limited cavity mode, a free-space buffer space of thousands of atoms~\cite{Manetsch2025Nature,Norcia2024} can be used to host the qubits that have already emitted photons while waiting for herald information, allowing for longer distance communication while saturating the rate upper bound by satisfying Eq.~\eqref{eq:continuous_operation}.

We now discuss the prospects for implementing quantum strategies for LCTC tasks with other leading quantum-network architectures.
The hardware requirements summarized in Table~\ref{table:hardware_requirements} are platform-agnostic, but different platforms realize different tradeoffs between local-cycle latency, telecom networking efficiency, and available memory depth for time-multiplexed operation.

Firstly, free-space neutral-atom arrays, without optical cavities, offer large coherent registers with coherence times ranging from seconds to over 10 seconds with a large qubit count~\cite{Manetsch2025Nature,Covey2023}. 
Telecom-band atom-photon networking with atom arrays has been demonstrated~\cite{CoveyNaturePhysics2025}, and entanglement generation over tens-of-kilometers of fiber has been demonstrated~\cite{vanLeent2022Nature}. 
Without the cavity enhancement, the HEG rates may be limited~\cite{young2022architecture}, calling for the design of substantial channel multiplexing, e.g.,~by scaling up the method demonstrated in Ref.~\cite{CoveyNaturePhysics2025} with a fiber array.

Secondly, trapped-ion systems provide comparable coherence and gate performance~\cite{helios2025}, and long-lived remote ion-ion entanglement has been demonstrated~\cite{Main2025Nature,Liu2026}. 
Rapid progress in photon-interfaced multi-ion registers~\cite{Canteri2025} and cavity- or fiber-integrated ion networking demonstrations~\cite{Krutyanskiy2023} points to a credible path toward high-rate HEG and short measurement cycles.

Solid-state spin defects combine long-lived spin memories with nanophotonic integration and have achieved metropolitan-scale entanglement distribution in telecom networks~\cite{Knaut2024Nature}. 
Single-shot readout in the 10--\qty{100}{\us} range has been demonstrated for both NV and SiV systems under cryogenic operation~\cite{Zhang2021NC,Sukachev2017PRL}, and continued improvements in photonic structures are steadily enhancing the photon collection efficiency.
While large per-node memory depth can be impeded by inhomogeneous spectral shifts and, for some defect species, the need for telecom conversion, recent progress on multiplexed multi-emitter nodes and frequency-multiplexed entanglement generation~\cite{Ruskuc2025} highlights concrete strategies to scale-up.

Finally, superconducting circuits offer fast local measurement (\qty{100}{\ns}--\qty{1}{\us}) with mature microwave control stacks~\cite{Kjaergaard2020AnnualReview}, and meter-scale remote entanglement with Bell-inequality violation has been demonstrated at high rates~\cite{Storz2023}. 
Coherence times have also continued to improve, with reports extending toward the millisecond regime in optimized devices~\cite{Somoroff2023PRL}. 
For these systems, the central obstacle to long-distance networking is the lack of a native telecom-photon interface, so practical deployment over fiber hinges on microwave-to-optical transduction with simultaneously high efficiency and low added noise.

Overall, while different physical platforms exhibit distinct performance characteristics, we have presented the hardware requirements summarized in Table~\ref{table:hardware_requirements} in a way that does not depend on the detailed mechanism of the memory system.
Feeding system-specific parameter values into the evaluation of $\tau_\text{dec}, \epsilon$ and $R_\text{HEG}(L)$ enables a concrete investigation of the applicability of the system for LCTC.

\section{Extension to multiparty tacit coordination with cQED network} \label{sec:multipartite}
\begin{figure*}
    \centering
    \includegraphics[width=0.98\linewidth]{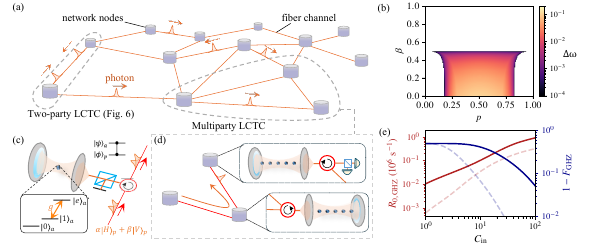}
    \caption{\textbf{Scheme for full network-level coordination.} 
    \textbf{(a)} LCTC tasks across quantum network. We illustrate the implementation of two-party LCTC (left, where remote atom-atom Bell pairs are generated using two-photon interference as shown in Fig.~\ref{fig:resource_estimation}) and three-party LCTC [right, which employs cavity-assisted photon scattering (CAPS) approach] as part of the full network. 
    \textbf{(b)} Simulation results of $k=3$ with balanced $\beta$ and the input
    distribution is independent Bernoulli distributed with probability $p$ for all three nodes.  
    The maximal gap matches the standard CHSH improvement $(\sqrt{2}-1)/4$.
    \textbf{(c, d)}
    Remote GHZ state generation using CAPS atom-photon gates. 
    (c) Here photon interacts with the atoms coupled to optical cavities via CAPS-based controlled-phase gate; (d) then following the sequential interaction with three cavity nodes, the photon is measured in the $X$ basis $\{(\ket{H}_p\pm \ket{V}_p)/\sqrt{2}\}$, thereby projecting the atomic states to three-qubit GHZ states up to local unitary.
    \textbf{(e)} 
    Simulated rate $R_{0,\mathrm{GHZ}}$ and infidelity $1-F_\text{GHZ}$ of remote GHZ states generated via CAPS atom-photon interaction as a function of internal cooperativity $C_\mathrm{in}$ for two photon pulse widths $\sigma_t/\gamma=0.12$ (solid) and $0.34$ (dashed), where we set $\gamma/2\pi = 0.24$ MHz. 
     }
    \label{fig:multipartite}
\end{figure*}

Typical classical coordination problems, such as those in finance, power grid systems, and communication networks discussed in Sec.~\ref{sec:application_examples}, already routinely require multiparty coordination with latency constraints, where decisions must be made without in-round classical communication, and the appropriate performance benchmark is the optimal no-signaling classical strategy. 
Quantum advantage in this setting requires surpassing that benchmark using preshared entanglement without relaxing the underlying latency constraint~\cite{2510.26349,Terzija2011,Arun2025}. 
As a concrete example, multi-venue trading subject to a single global inventory or risk constraint can be formalized as a $k$-party XOR game for $k>2$, in which each party observes a local binary signal, and the utility depends on the parity of outputs~\cite{2510.26349}. 
In such tasks, multipartite entanglement enables access to genuine $k$-party correlations that can yield a provable advantage.
Combining multiple instances of multiparty coordination would enable LCTC over a more complex network of parties [Fig.~\ref{fig:multipartite}(a)].
As such, here we extend our framework, developed for two-party LCTC in Secs.~\ref{sec:problem_statement}--\ref{sec:resource_estimation}, to multiparty LCTC and analyze its concrete implementation using a high-performance cavity-assisted multipartite entanglement generation protocol.

\subsection{Multiparty XOR game} \label{sec:multipartite_protocol}

For multiparty LCTC tasks, we consider $k> 2$ parties indexed by $i\in\{1,\dots,k\}$, and in each round, party $i$ receives a binary input $x_i\in\{0,1\}$ and outputs a binary decision $a_i\in\{0,1\}$. The input tuple $\bm{x}=(x_1, x_2,\dots,x_k)$ is drawn from a known distribution $P(\bm{x})$. The LC condition~\eqref{eq:lc_condition} is enforced by requiring that the outputs must be generated from pre-established resources only (shared randomness for classical strategies or preshared entanglement for quantum strategies), without signaling among parties during the local decision window.

As an illustrative example, we consider a three-party $(k=3)$ XOR game in which the utility function $u(\bm{a}\mid \bm{x})$ is partitioned according to the output parity $a_\oplus = \bigoplus_{i=1}^k a_i$ by defining $u(o\mid\bm{x}) = u(\bm{a}\mid \bm{x})$ for any output tuple $\bm{a} = (a_1,a_2, \cdots, a_k)$ satisfying $a_\oplus = o$.
We first consider the uniform input distribution $P(\bm{x}) = 1/8$ and that the utility is determined by the majority of the input values,
\begin{equation}
    \bar{u}(o\mid \bm{x}) = 
    \begin{cases}
        1- \Maj(\bm{x}), & o= 0, \\
        \Maj(\bm{x}), & o=1,
    \end{cases}
\end{equation}
where
\begin{equation}
    \Maj(\bm{x}) = 
    \begin{cases}
        1, & \sum_{i} x_i > 3/2, \\
        0, & \text{otherwise}.
    \end{cases}
\end{equation}
In this case, the expected utility 
\begin{equation}
    \omega = \sum_{\bm{x}}P(\bm{x}) \sum_{\bm{a}} u(\bm{a}\mid \bm{x}) P(\bm{a}\mid \bm{x})
\end{equation}
reduces to the winning probability of the game, where any classical strategy cannot exceed $\omega_\text{C} = 3/4$ while the optimal quantum strategy with a Greenberger-Horne-Zeilinger (GHZ) state $\ket{\text{GHZ}} = (\ket{000}+ \ket{111})/\sqrt{2} $ achieves $\omega_\text{Q} = (1+1/\sqrt{2})/2$~\cite{Svetlichny1987, Mitchell2004, Brunner2014RMP}, as in the CHSH game in Eq.~\eqref{eq:chsh_predicate} (see Appendix~\ref{app:multipartite} for the details).

To reflect the confidence of the majority vote in the utility, we then classify the input $\bm{x}$ according to whether its Hamming weight $\wt(\bm{x}) = \sum_{i} x_i$ satisfies $\wt(\bm{x}) = 0 ~\text{mod}~3$, and we introduce a parameter $\beta \in [0,1]$ as
\begin{equation} \label{eq:three_party_u(o|x)}
    \begin{aligned}
        &u(o \mid \bm{x})  \\
        &=
        \begin{cases}
            \bar{u}(o\mid \bm{x}) , & \wt(\bm{x}) = 0 ~ \text{mod}~ 3, \\
            (1-\beta)\bar{u}(o\mid \bm{x}) + \beta [1-\bar{u}(o\mid \bm{x})] , & \text{otherwise}.
        \end{cases}
    \end{aligned}
\end{equation}
In a quantum strategy, as in the two-party cases in Sec.~\ref{sec:extended_protocols}, each party measures a $\pm1$-valued observable.
In this case, a GHZ state $\ket{\text{GHZ}}$ allows the optimal strategy, leading to the quantum value as~\cite{Werner2001} 
\begin{equation} \label{eq:omega_Q_three_XOR}
    \omega_\text{Q} = \frac{1}{2} \ab[ 1+ \max_{\{\varphi_0,\cdots,\varphi_3\}} \sum_{\bm{x}}M_{\bm{x}} \cos\ab(\varphi_0 + \sum_{i=1}^{3}\varphi_i x_i)],
\end{equation}
where 
\begin{equation}
    M_{\bm{x}} = P(\bm{x}) \sum_{o\in \{0,1\}}(-1)^{o} u(o\mid \bm{x}),
\end{equation}
and $\{\varphi_0, \cdots, \varphi_3\}$ characterizes the measurement basis of three parties (see Appendix~\ref{app:multipartite} for the details).
Fig.~\ref{fig:multipartite}(b) shows the gap $\Delta \omega$, where each input is sampled from the Bernoulli distribution with probability $p$ and $P(\bm{x}) = p^{\wt(\bm{x})}(1-p)^{3-\wt(\bm{x})}$.
We find that the quantum advantage $\Delta \omega > \num{e-4}$ is available for $\beta < 1/2$ and a wide range of $p$.

Furthermore, the operational criteria and the hardware requirements, as summarized in Table~\ref{table:hardware_requirements}, apply to multiparty coordination in the same way as in the two-party case: each coordination round consumes preshared multipartite entangled qubits and must complete local basis selection and measurement within the local decision window, while multiplexed memories absorb link latency into a throughput requirement.
In the following, we discuss a realistic scheme to generate multipartite entangled states across the network with cQED network nodes, completing the proposal for network-level LCTC.

\subsection{GHZ state generation with cQED network nodes}

In a cQED system, a single photon can interact directly with atomic qubits via scattering from a one-sided cavity, which imprints state-dependent phases onto the composite atom-photon state~\cite{Duan2004,Reiserer2015RMP,Reiserer2013Science,Tiecke2014Nature,2507.01229} [Fig.~\ref{fig:multipartite}(c); see also Appendix~\ref{app:cQED_physical_model}].
Such a cavity-assisted photon scattering (CAPS) enables a high-fidelity, high-success-probability implementation of atom-photon controlled-phase (CZ) gates, leading to a high entanglement generation rate and fidelity~\cite{2507.01229}.
CAPS can be utilized to efficiently generate a GHZ state across the network with three or more nodes, as illustrated in Fig.~\ref{fig:multipartite}(d):
initially, the atoms are prepared in $\ket{+}=(\ket{0}+\ket{1})/\sqrt{2}$, and a polarization photonic qubit $\ket{+}_p = (\ket{H}_p + \ket{V}_p)/\sqrt{2}$, where $\ket{H(V)}_p$ represents a $H(V)$-polarized photon, is launched into a network of circulators and atom-cavity systems.  
After the photon interacts with all three nodes, a photonic qubit measurement heralds the successful generation of the GHZ state up to local Pauli gates (see the detailed protocol in Appendix~\ref{app:cQED_physical_model}).

To see the concrete rate-fidelity performance of the CAPS-based GHZ state generation protocol, we perform a numerical evaluation of the atom-cavity dynamics for $k=3$ and identify the realistic GHZ state generation rate and fidelity;
we set all three nodes to have equal cooperativity $C_{\rm in}$ and treat the driving photon as a Gaussian-shaped pulse with a pulse duration (amplitude envelope standard deviation) of $\sigma_t$; here, for a realistic evaluation, we assume that the photon source is also considered to be imperfect, being generated by the same atom-cavity system as the other network nodes with non-unity emission probability~\cite{Utsugi2022}.  
Following the numerical and analytical modeling of the error mechanisms (see Appendix~\ref{app:cQED_physical_model} and Ref.~\cite{2507.01229}), we compute both the heralded entanglement generation rate $R_{0,\mathrm{GHZ}}$ and the fidelity $F_\text{GHZ}$, which are plotted in Fig.~\ref{fig:multipartite}(e) for varying $C_\mathrm{in}$, with $\sigma_t$ set to $\sigma_t/\gamma = 0.12$ (solid). 
While short photon pulses achieve faster repetition rates and higher $R_\mathrm{0, GHZ}$, the spectrum of the photon is broader, and the infidelity $1-F_\text{GHZ}$ is high, particularly for low $C_\mathrm{in}$.
Therefore, we also plot the results for $\sigma_t/\gamma = 0.34$ (dashed), which demonstrate higher fidelity at the cost of the generation rate.
With sufficiently high $C_\mathrm{in}$, the entanglement generation rates reach close to $10^6~\text{s}^{-1}$, with GHZ-state infidelity close to 0.05; assuming $\epsilon_\text{meas} = 0.01$, this achieves the combined-infidelity comfortably below its threshold of $1-1/\sqrt{2}$ for $\beta = 0$ and a uniform input distribution (see Appendix~\ref{app:multipartite} for details).

The simulated performance in Fig.~\ref{fig:multipartite}(e) indicates that CAPS-based remote GHZ generation can reach high HEG rates while maintaining high fidelity over the cooperativity range relevant to near-term atom-cavity devices. At these fidelities, the quantum-classical gap in Fig.~\ref{fig:multipartite}(b) will remain positive for the benchmark case, i.e., $\beta = 0$.
Under the positive-gap regime, the required rate $R_\text{req}(\alpha)$ is evaluated in the same way as the two-party case; for the utility function in Eq.~\eqref{eq:three_party_u(o|x)}, the $p$-value is given by Eq.~\eqref{eq:pvalue_binomial}, and then $R_\text{req}(\alpha) = n_\text{req}(\alpha)/T_\text{env}$ with $n_\text{req}(\alpha)$ in Eq.~\eqref{eq:n(alpha)_def} gives the required GHZ state generation rate.
The simulated high-rate remote GHZ state generation protocol will therefore demonstrate a prospect for certifiable quantum advantage in multiparty LCTC, which is to be supported by the implementation of time-multiplexed architecture as in the two-party case illustrated in Fig.~\ref{fig:multiplexing}; Ref.~\cite{2507.01229} analyzes such time-multiplexed CAPS operations, providing a baseline for more comprehensive evaluations.

\section{Discussion}\label{sec:discussions_outlook}

In this work, we thoroughly analyzed the required quantum resources to achieve robust quantum advantage in LCTC tasks.
First, we clarified the mapping between nonlocal games and LCTC tasks, which resulted in the incorporation of a natural notion of a finite environment stationary window.
Such an analysis allowed us to quantify the \emph{operational criteria} for quantum advantage, including the required rate and fidelity of remote entanglement generation for a statistically significant quantum advantage, which is substantially more stringent than the previous one considered in Ref.~\cite{2407.21723}.

To implement such high-performance quantum network operations, we proposed a multiplexed, memory-based network architecture, showing how buffering and time-multiplexed scheduling achieve the required rate;
a more concrete hardware proposal is provided based on trapped ${}^{171}$Yb atoms in optical cavities, combining both high-rate, direct telecom-band networking and exquisite qubit control recently proposed and demonstrated.
We have further extended the analysis to the multipartite case, proposing both a nonlocal game formulation for general multiparty coordination tasks and a blueprint for generating multipartite entangled states across multiple locations at high rates and fidelity using the atom-cavity system proposed.
Based on a thorough analysis of atom-cavity dynamics and realistic overhead estimates for both bipartite and three-party cases, it is expected that the proposed atom-cavity systems, with modest performance of $C_\mathrm{in}=20$, can achieve the entanglement generation rate and fidelity that satisfy the demands of highly relevant LCTC tasks.
These results provide a clear path towards near-term demonstrations of certified quantum advantage in time-critical distributed coordination tasks.

We emphasize that our framework is designed to model the LCTC tasks and is not intended for the continued effort on the rigorous testing of LHV models~\cite{Hensen2015Nature,Rosenfeld2017} and the development of robust cryptographic protocols~\cite{Zapatero2023npjQI, Acin2007}.
As such, while some mathematical formulations and analysis techniques are similar, several foundational analyses for the Bell tests may not directly apply to the analysis of quantum advantage in LCTC tasks, such as \emph{loophole} analyses and the strictness of the no-signaling condition~\cite{Brunner2014RMP,Hensen2015Nature,Storz2023,Zeng2025Quantum}. 
For example, classical communication latencies between the parties may incorporate additional constraints for communication than those considered for the tests of LHV models; in addition to the relativistic limit, classical communication latencies arising from regulatory limits and classical preprocessing overhead are also relevant when evaluating the realistic quantum advantage in LCTC tasks.

At the same time, the model we analyze remains deliberately simplified to stay analytically and numerically tractable: inputs and actions are discrete, rounds are treated as conditionally independent, and an input distribution and a utility function are assumed to be fixed over the environment stationary window. 
Real-world LCTC tasks would exhibit continuous-variable and nonstationary signals, correlated rounds, heterogeneous constraints across parties, and additional objectives that cannot be captured by a single static utility table. 
Extending the framework to richer action spaces and sequential policies under drift and model mismatch, while preserving a principled finite-statistics guarantee, is an important direction for completing the link between the theoretical analysis of quantum advantage and deployable coordinated decision-making systems.

Finally, while this work focuses on nonlocal games and LCTC tasks, distributed quantum processors can exhibit quantum advantages in other aspects, such as energy-consumption advantage~\cite{PRXEnergy.4.023008}, communication complexity reduction~\cite{2411.03240LOCAL,Tang2023BQC}, privacy-preserving delegated computation~\cite{Li2024BQML}, networked sensing and metrology~\cite{Stas2026,Yang2024PRL}, as well as the scalable implementation of fault-tolerant quantum computers~\cite{Sunami2025PRXQ,sunami2025transversalsurfacecodegamepowered,sunami2025boosting,mohseni2025buildquantumsupercomputerscaling,Sinclair2025PRR}.
The proposed quantum network nodes are well-suited as a common platform for these protocols because they combine 
large-scale qubit array, high-fidelity qubit control, long-lived memories, and fiber-native photonic links.
Crucially, it is expected that a wider variety of robust and practically relevant quantum advantages will be realizable with such a quantum network supplemented by the capability of local quantum operations, which represent an attractive near-term target for quantum technology development and deployment before the realization of large-scale, fault-tolerant quantum computers.

\bibliography{refs}

\clearpage
\appendix

\section*{Appendices}
Appendices are organized as follows.
In Sec.~\ref{app:opt_measurement_angle}, we describe the numerical optimization routine for the generalized LCTC tasks in Sec.~\ref{sec:operational_criteria}.
In Sec.~\ref{app:multipartite}, we provide details of the multiparty XOR game discussed in Sec.~\ref{sec:multipartite}, including the analysis of the infidelity threshold.
Finally, in Sec.~\ref{app:cQED_physical_model}, we provide the theoretical model used for the results in Fig.~\ref{fig:resource_estimation} and Fig.~\ref{fig:multipartite}, where we analyze the rate and fidelity of remote Bell pair and GHZ state generation, as well as the cavity-assisted measurement protocol.

\section{Optimal measurement angles for XOR games} \label{app:opt_measurement_angle}

\begin{figure}[t]
    \centering
    \includegraphics[width=\linewidth]{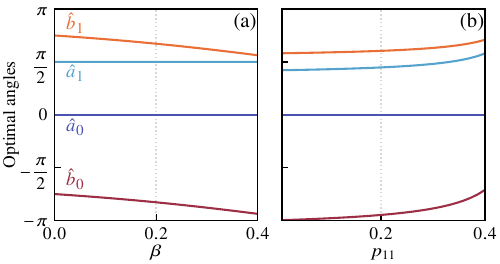}
    \caption{Optimal measurement angles for the generalized LCTC tasks. We employ the nonlocal game in Fig.~\ref{fig:extended_protocols}(a) with $p = 1/2$ for panel (a), and one in Fig.~\ref{fig:extended_protocols}(b) with $\beta=0$ for panel (b).}
    \label{fig:optimal_meas_angle}
\end{figure}

Here, we show the optimal measurement observables for Alice and Bob in the generalized LCTC tasks discussed in Sec.~\ref{sec:operational_criteria}.
Since $Q(M)$ in Eq.~\eqref{eq:Q(M)} is characterized by the angle between two unit vectors, $\hat{a}_x$ and $\hat{b}_y$, it is sufficient to search for two vectors in the $xy$ plane; we fix $\hat{a}_{0} = (1,0,0)$ and set the other three vectors as $\hat{a}_1 = (\cos\theta,\sin\theta,0)$, and $\hat{b}_i = (\cos\phi_i,\sin\phi_i,0)$. We then numerically optimize the angles to achieve the optimal value.
Figs.~\ref{fig:optimal_meas_angle}(a) and (b) present the numerical results for the two models shown in Figs.~\ref{fig:extended_protocols}(a) and (b), respectively. For the model with $\beta_1= \beta_2 = \beta$ and $P(x,y)=1/4$ [Fig.~\ref{fig:optimal_meas_angle}(a)], the results with $\beta=0$ reproduce the well-known results, $(\theta,\phi_1,\phi_2) = (\pi/2, 3\pi/4, -3\pi/4)$ for the CHSH game~\cite{Brunner2014RMP}.
In contrast, since the off-diagonal terms of $M$ decrease as $\beta$ increases, Bob's measurement angles approach parallel to Alice's angles so that the parties prioritize maximizing $-M_{0,0}\hat{a}_0\cdot \hat{b}_0 -M_{1,1}\hat{a}_1\cdot \hat{b}_1$.
While the nonlocal game with uniform inputs on the two parties leads to fixed angles for Alice's measurement 
[Fig.~\ref{fig:optimal_meas_angle}(a)], the correlated-input case, shown in Fig.~\ref{fig:optimal_meas_angle}(b), changes both measurement angles as the input distribution changes.
These numerical results explicitly provide the optimal and feasible quantum strategy to achieve the largest gap.

\section{Details of analysis of multiparty LCTC tasks} \label{app:multipartite}

Here we summarize the model and numerical procedure used to evaluate the $k$-party LCTC task for $k>2$ shown in Sec.~\ref{sec:multipartite}. 
In Sec.~\ref{app:multi-party-xor}, we first provide a brief overview of the expected utility in the multiparty XOR game and provide a more detailed account for the $k=3$ case in Sec.~\ref{app:three-party-xor}.
In Sec.~\ref{app:infid-threshold-k3}, we further discuss the infidelity threshold for the $k=3$ case.

\subsection{Multiparty XOR game}
\label{app:multi-party-xor}
We consider binary inputs $x_i\in\{0,1\}$ and binary outputs $a_i\in\{0,1\}$ for $i\in\{1,\dots,k\}$, with an arbitrary full-support input distribution $P(\bm{x})$, where $\bm{x} = (x_1,x_2,\cdots,x_k)$. For a fixed conditional probability distribution $P(\bm{a}|\bm{x})$, where $\bm{a} = (a_1,\dots,a_k)$, the expected utility is
\begin{equation}
    \omega = \sum_{\bm{x}} P(\bm{x}) \sum_{\bm{a}} u(\bm{a}\mid \bm{x}) P(\bm{a} \mid \bm{x}) 
\end{equation}
for a given utility function $u(\bm{a}|\bm{x})$.
In the following, we assume no-signaling constraints between any pairs of parties and consider that a classical strategy only has a local correlation, where the probability distribution can be written in the form~\cite{Brunner2014RMP}
\begin{equation}
    P(\bm{a}\mid \bm{x}) = \int \dd{\lambda}P(\lambda) \bigoplus_{i=1}^k P_i(a_i\mid x_i, \lambda),
\end{equation}
where $\lambda$ represents a shared local random variable.

\subsection{Three-party XOR game}
\label{app:three-party-xor}

Here, for $k = 3$, we consider an XOR game: a utility function is classified according to the output parity $a_\oplus \equiv \bigoplus_{i=1}^k a_i$ by defining $u(o \mid \bm{x}) = u(\bm{a}|\bm{x})$ for any $\bm{a}$ satisfying $a_\oplus = o$.
As in the two-party XOR game in Sec.~\ref{sec:non-local_games}, we find
\begin{equation}
    \sum_{\bm{a}} u(\bm{a}\mid \bm{x}) P(\bm{a} \mid \bm{x}) = \sum_{o \in \{0,1\}} u(o \mid \bm{x}) \frac{1 + (-1)^o E_{\bm{x}}}{2},
\end{equation}
where 
\begin{equation}
    E_{\bm{x}} = \sum_{\bm{a}} (-1)^{a_\oplus} P(\bm{a}\mid \bm{x}).
\end{equation}

As a concrete nonlocal game, we first consider the uniform input distribution $P(\bm{x}) = 1/8$ and that the majority of the input values determines the utility, i.e.,
\begin{equation}
    \bar{u}(o\mid \bm{x}) = 
    \begin{cases}
        1- \Maj(\bm{x}), & o= 0, \\
        \Maj(\bm{x}), & o=1,
    \end{cases}
\end{equation}
where
\begin{equation}
    \Maj(\bm{x}) = 
    \begin{cases}
        1, & \sum_{i} x_i > 3/2, \\
        0, & \text{otherwise}.
    \end{cases}
\end{equation}
In this case, we find
\begin{equation}
    \sum_{\bm{a}} u(\bm{a}\mid \bm{x}) P(\bm{a} \mid \bm{x}) = \frac{1}{2} \ab[1 + (-1)^{\Maj(\bm{x})} E_{\bm{x}}],
\end{equation}
leading to
\begin{equation}
    \omega = \frac{1}{2}\ab(1 + \frac{S_3}{8}),
\end{equation}
with
\begin{equation}
    S_3 = \sum_{\bm{x}} (-1)^{\Maj(\bm{x})}E_{\bm{x}}.
\end{equation}
From the Svetlichny's inequality~\cite{Svetlichny1987,Brunner2014RMP}, any classical strategy cannot achieve $S_3 > 4$, leading to $\omega_\text{C} = 0.75$.
In the meantime, to maximize the expected utility with quantum strategies, it is sufficient that three parties utilize a GHZ state $\ket{\text{GHZ}} = (\ket{000} + \ket{111})/\sqrt{2}$ and measure their qubits in the $xy$ plane~\cite{Werner2001}; the measurement axis $\hat{a}_{x_i}^{i}$ for party $i \in \{1,2,3\}$ is given by 
\begin{equation}
\label{eq:measurement_basis_multiparty}
    \begin{aligned}
        \hat{a}_{0}^{i} =& (\cos(\varphi_0/3), \sin(\varphi_0/3),0), \\
        \hat{a}_{1}^{i} =& (\cos(\varphi_0/3 + \varphi_i), \sin(\varphi_0/3 + \varphi_i),0),
    \end{aligned}
\end{equation}
where $\{\varphi_0, \cdots, \varphi_3\}$ is a set of measurement angles. This yields
\begin{equation} \label{eq:E_x_multiparty}
    E_{\bm{x}} = \cos\ab(\varphi_0 + \sum_{i=1}^{3}\varphi_i x_i).
\end{equation}
By setting $\varphi_0 = -\pi/4$ and $\varphi_1 = \varphi_2 = \varphi_3 = \pi/2$, the quantum strategy achieves $S_3 = 4\sqrt{2}$~\cite{Mitchell2004}, leading to $\omega_\text{Q} = (1+1/\sqrt{2})/2$.

Here, as discussed in Sec.~\ref{sec:multipartite}, we employ a utility function that incorporates a softness parameter $\beta \in [0,1]$:
\begin{equation}
    \begin{aligned}
        &u(o \mid \bm{x})  \\
        &=
        \begin{cases}
            \bar{u}(o\mid \bm{x}) , & \wt(\bm{x}) = 0 ~ \text{mod}~ 3, \\
            (1-\beta)\bar{u}(o\mid \bm{x}) + \beta [1-\bar{u}(o\mid \bm{x})] , & \text{otherwise},
        \end{cases}
    \end{aligned}
\end{equation}
where $\wt(\bm{x}) = \sum_{i} x_i$ represents the Hamming weight of an input.
This leads to 
\begin{equation}
    \omega = \sum_{\bm{x}} P(\bm{x}) \sum_{\bm{a}} u(\bm{a}\mid \bm{x}) P(\bm{a} \mid \bm{x}) = \frac{1}{2}\ab(1 + \sum_{\bm{x}} M_{\bm{x}} E_{\bm{x}}),
\end{equation}
with 
\begin{equation}
    M_{\bm{x}} = P(\bm{x}) \sum_{o\in \{0,1\}} (-1)^o u(o \mid \bm{x}).
\end{equation}

For a classical strategy, we simulate $2^{2\times 3} = 64$ deterministic strategies and obtain the classical value $\omega_\text{C}$ as the maximum.
For a quantum strategy, from Eq.~\eqref{eq:E_x_multiparty}, the quantum value is calculated as
\begin{equation} \label{app_eq:omega_Q_GHZ}
    \omega_\text{Q} = \frac{1}{2} \ab[ 1+ \max_{\{\varphi_0,\cdots,\varphi_3\}} \sum_{\bm{x}}M_{\bm{x}} \cos\ab(\varphi_0 + \sum_{i=1}^{3}\varphi_i x_i)].
\end{equation}

\subsection{Infidelity threshold for $k=3$}
\label{app:infid-threshold-k3}

Here, we consider an error model as in the two-party case in Eq.~\eqref{eq:werner}; the erroneous state is given as
\begin{equation}
    \begin{aligned}
        &\rho(\epsilon_\text{GHZ})\\
        &= (1-\epsilon_\text{GHZ}) \ketbra{\text{GHZ}}{\text{GHZ}} + \frac{\epsilon_\text{GHZ}}{7} \ab(\mathbb{I}^3 - \ketbra{\text{GHZ}}{\text{GHZ}}) \\
        &= \ab(1-\frac{8\epsilon_\text{GHZ}}{7}) \ketbra{\text{GHZ}}{\text{GHZ}} + \frac{\epsilon_\text{GHZ}}{7} \mathbb{I}^3.
    \end{aligned}
\end{equation}
By using the error model for qubit measurement in Eq.~\eqref{eq:A_x(epsilon)}, we then find
\begin{equation}
    E_{\bm{x}}(\epsilon_\text{GHZ},\epsilon_\text{meas}) = (1-\epsilon^\prime) \cos\ab(\varphi_0 + \sum_{i=1}^{3}\varphi_i x_i),
\end{equation}
where the combined infidelity is given by
\begin{equation}
    \epsilon^\prime = 1- \ab(1-\frac{8\epsilon_\text{GHZ}}{7}) \ab(1-2\epsilon_\text{meas})^3.
\end{equation}
The state gives the quantum value with combined infidelity $\epsilon^\prime$ as
\begin{equation}
    \omega_\text{Q}(\epsilon^\prime) = \omega_\text{Q}(\epsilon^\prime = 0) - \epsilon^\prime \ab[\omega_\text{Q}(\epsilon^\prime = 0) -\frac{1}{2}],
\end{equation}
where $\omega_\text{Q}(\epsilon^\prime =0)$ reduces to $\omega_\text{Q}$ in Eq.~\eqref{app_eq:omega_Q_GHZ}.
Thus, the quantum-classical gap is given by
\begin{equation}
    \Delta \omega(\epsilon^\prime) = \Delta \omega(\epsilon^\prime=0) -\epsilon^\prime \ab[\omega_\text{Q}(\epsilon^\prime=0) -\frac{1}{2}],
\end{equation} 
showing that the advantage $\Delta \omega(\epsilon^\prime) > 0$ requires
\begin{equation}
    \epsilon^\prime < \frac{\Delta \omega(\epsilon^\prime = 0)}{\omega_\text{Q}(\epsilon^\prime=0) -1/2},
\end{equation}
under the assumption $\omega_\text{Q}(\epsilon^\prime=0) > 1/2$.
Considering $\beta = 0$ and $P(\bm{x}) = 1/8$ as a representative case, the inequality is given by
\begin{equation}
    \epsilon^\prime < 1-\frac{1}{\sqrt{2}},
\end{equation}
exhibiting the same threshold for the CHSH game in Eq.~\eqref{eq:Delta_omega_epsilon_CHSH}.

\section{Implementation of nonlocal games with cavity-QED systems}  \label{app:cQED_physical_model}

Here, we discuss the details of physical protocols to implement remote entanglement generation and qubit measurements with cavity-QED (cQED) systems described in Sec.~\ref{sec:resource_estimation} and Sec.~\ref{sec:multipartite}. 
In Sec.~\ref{app:Bell_state_TPI}, we present the protocol and the detailed error model for remote HEG by two-photon interference. 
In Sec.~\ref{app:cavity_readout}, we then discuss cavity-enhanced qubit measurement and its simulation details presented in Fig.~\ref{fig:resource_estimation}(d).
We finally give the detailed protocol of the remote GHZ-state generation with cavity-assisted photon scattering, utilized for multiparty nonlocal games discussed in Sec.~\ref{sec:multipartite}.

\subsection{Bell states generated by two-photon interference} \label{app:Bell_state_TPI}

Here, we give an error model of the Bell state generated with the two-photon interference (TPI); Alice and Bob host identical systems, and each generates an atom-photon Bell state $\ket{\Phi^{+}} = (\ket{00} + \ket{11})/\sqrt{2}$ through photon generation, where the photonic qubit is encoded by time bins.
The two emitted photons are sent to the Bell-measurement apparatus at the midpoint, resulting in a remote atom-atom Bell state with a finite probability.
To incorporate errors arising from reexcitation~\cite{Krutyanskiy2023}, we consider that two parties generate atom-photon entanglement where the photon is emitted in a classical mixture of the temporal modes $v_l(t)\,(l\in \mathbb{N}_+)$
with probabilities $p_l$.
The photon temporal distribution is characterized by the autocorrelation function~\cite{Fabre2020},
\begin{equation}
    g^{(1)}(t,t^\prime) = \sum_l p_l v_l^\ast(t)v_l(t^\prime),
\end{equation}
leading to the generation probability $p_e = \int \dd{t} g^{(1)}(t, t)$.

For two clicks at times $(t_0,t_1)$ in a  pattern of detectors that announces the success of the photonic Bell measurement, the two atoms are projected onto~\cite{2507.01229}
\begin{widetext}
\begin{equation}
    \rho(t_0,t_1) = \frac{1}{16 p(t_0,t_1)}\mqty{g^{(1)}(t_0,t_0)g^{(1)}(t_1,t_1) & -\vab{g^{(1)}(t_0,t_1)}^2 \\ -\vab{g^{(1)}(t_0,t_1)}^2 & g^{(1)}(t_0,t_0)g^{(1)}(t_1,t_1)}
\end{equation}
\end{widetext}
in the basis of $\{\ket{0}\ket{1}, \ket{1}\ket{0}\}$, where $p(t_0,t_1) = g^{(1)}(t_0,t_0)g^{(1)}(t_1,t_1)/8$ represents the detection probability.
This results in the detection-time-averaged state as 
\begin{equation}
    \begin{aligned}
        \rho =& \frac{\iint \odif{t_0}\odif{t_1} p(t_0,t_1) \rho(t_0,t_1)}{\iint \odif{t_0}\odif{t_1}p(t_0,t_1)} = \frac{1}{2}\mqty{1 & V \\ V& 1} \\
        =& \frac{1+V}{2}\ketbra{\Psi^{-}}{\Psi^{-}} + \frac{1-V}{2}\ketbra{\Psi^{+}}{\Psi^{+}},
    \end{aligned}
    \label{eq:phi-mix}
\end{equation}
where $V$ represents a single-photon trace purity~\cite{Fischer2018, Trivedi2020, Kikura2025_high_purity}, i.e.,
\begin{equation}
    V = \frac{\iint \mathrm{d}t\dd{t^\prime} |g^{(1)}(t,t^\prime)|^2}{\ab[\int \dd{t} g^{(1)}(t,t)]^2}.
    \label{eq:mode_overlap_functional}
\end{equation}

In our proposal in Sec.~\ref{sec:resource_estimation}, we utilize the metastable qubit in \text{Yb} atoms to generate remote atom-atom Bell pairs. 
See the detailed protocol in the caption of Fig.~\ref{fig:resource_estimation}.
Here, we incorporate the reexcitation as an infidelity source; the atom spontaneously decays to the initial state at rate $\gamma_{^{3}\text{D}_1}/2\pi = \qty{0.24}{\MHz}$ and the decayed state may be reexcited by the excitation pulse of Rabi frequency $\Omega/2\pi = \qty{30}{\MHz}$~\cite{CoveyNaturePhysics2025}.
We numerically calculate performance metrics of TPI-based HEG (see Ref.~\cite{2507.01229} for the details) and present the rate $R_0 = p_e^2/(2\tau_e)$ in Fig.~\ref{fig:resource_estimation}(b).
We use $g/2\pi = \qty{3}{\MHz}$, and $\kappa_\text{ex} = g + 2\kappa_\text{in}$ with $\kappa_\text{in} = g^2/(2\gamma_{^{3}\text{D}_1} C_\text{in})$, ensuring the infidelity $(1-V)/2 < 0.01$ for $C_\text{in} \in [1,100]$.

\subsection{Cavity-enhanced qubit measurement}
\label{app:cavity_readout}

To study the performance of cavity-based local measurement, we perform numerical simulations on the error model in a previous study that implemented cavity-enhanced fluorescence measurement~\cite{Deist2022PRL}. 
An atomic state coupled to the cavity, called a bright state, repeatedly emits photons through a probe duration $\tau_\text{meas}$, collected by a detector at rate $R_\text{bright}$.
In contrast, the uncoupled state is supposed not to emit photons; the contrast allows us to measure the qubit with high fidelity.
In a realistic situation, however, the dark state also clicks the detector at rate $R_\text{dark}$
due to experimental imperfection; in our case, a finite-amplitude vector ac Stark shift and/or bias magnetic field for decoupling the dark state from the cavity might cause a small fraction of photon emission.
Thus, we optimize the threshold $n_\text{th} \in \{0,1,\cdots\}$ 
of the photon-detection number to minimize the measurement infidelity; we judge a bright state in the case of the measured photon number being larger than $n_\text{th}$ and otherwise judge a dark state.
For simulation simplicity, we assume that the detected photon distribution follows a Poisson distribution.
The false-positive probability is given by
\begin{equation}
    P^{+} = 1 - \bar{Q}(n_\text{th}, R_\text{dark}\tau_\text{meas}),
\end{equation}
with the cumulative distribution function of a Poisson distribution, $\bar{Q}(n,\lambda) = e^{-\lambda} \sum_{m=0}^{n} \lambda^m/m!$.
In contrast, for the bright state, we incorporate the effect induced by blowing out the atom from the cyclic transition, leading to the false-negative probability as 
\begin{equation}
    P^{-} = \int_0^{\tau_\text{meas}} \dd{t} \rho_t(t) \bar{Q}(n_\text{th},Rt),
\end{equation}
where $\rho_t(t) = e^{-t/T_\text{life}}/T_\text{life} + e^{-\tau_\text{meas}/T_\text{life}}\delta(t-\tau)$ with the bright-state lifetime $T_\text{life}$ through the probe.
While Ref.~\cite{Deist2022PRL} collected photons from both qubit states by inserting a pulse that transfers an atom in a dark state to a bright one to detect the atom loss, we skip the latter probe to reduce the measurement time by a factor of $\sim 1/2$.
In this case, the measurement infidelity, i.e., the misclassification probability, is given by $P^{+} + P^{-}$.

For cavity-enhanced measurement, the bright-state photon-collection rate is given by $R_{\mathrm{bright}} = \eta_{\mathrm{det}}\, \kappa_{\mathrm{ex}} g^2 /(4\kappa^2)$, where $\eta_{\mathrm{det}}$ denotes the detector efficiency~\cite{Deist2022PRL}. The dark-state off-resonant scattering rate is suppressed by the detuning $\Delta$ and scales as $R_{\mathrm{dark}} \sim R_{\mathrm{bright}} (\gamma_{^3\mathrm{P}_1}/\Delta)^2$. Note that the effect arising from dark counts of commercial detectors, typically at the level of 1--\qty{100}{\Hz}, are negligible for the extracted infidelity on the microsecond measurement timescales considered here. 
In Fig.~\ref{fig:resource_estimation}(d), we use $g/2\pi= \qty{3}{\MHz}$, $\kappa_\text{ex} = (g + 2\kappa_\text{in})/3$, $\gamma_{^3\mathrm{P}_1}/2\pi =\qty{91}{\kHz}$, $\kappa_\text{in} = g^2/(2 \gamma_{^3\mathrm{P}_1}C_\text{in})$, $T_{\mathrm{life}} = \qty{1.6}{\ms}$, and $\Delta/\gamma_{^3\mathrm{P}_1} \gtrsim 100$, for which the contribution of $R_{\mathrm{dark}}$ is sufficiently small.

\subsection{GHZ-state generation with CAPS} \label{subapp:GHZ_generation}

\begin{figure}[t]
    \centering
    \includegraphics[width=\linewidth]{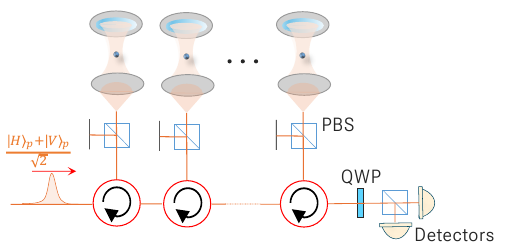}
    \caption{Schematic for the heralded generation of Greenberger-Horne-Zeilinger states shared by multiple parties.
    At each party, we apply the atom-photon controlled-phase gate with a polarizing beamsplitter (PBS). The photon is finally measured in the basis $\{(\ket{H}_p \pm \ket{V}_p)/\sqrt{2}\}$ with a quarter-wave plate (QWP), a PBS, and photon detectors.}
    \label{fig:GHZ_diagram}
\end{figure}

Here we present the protocol of remote GHZ-state generation, similar to one presented in Ref.~\cite{Zheng2012}.
Without loss of generality, we consider three parties labeled by $j\in\{1,2,3\}$, while the protocol can be extended for more than three parties.
Each party initially prepares an atomic qubit as $\ket{+}_{a,j}$ in an optical cavity.
As shown in Fig.~\ref{fig:GHZ_diagram}, we then prepare a single photon as $(\ket{H}_p + \ket{V}_p)/\sqrt{2}$ and route it to the detection apparatus through three atom-cavity systems; the photon $\ket{V}_p$ sequentially reflects off the cavities hosted by each party, thereby applying the phase gate to each atom.
In the ideal case, this results in the photon-atom entangled state $(\ket{H}_p \ket{+++}_a - \ket{V}_p \ket{---}_a)/\sqrt{2}$, where $\ket{\mu_1\mu_2\mu_3}_a = \ket{\mu_1}_{a,1}\ket{\mu_2}_{a,2}\ket{\mu_3}_{a,3}$; the photonic measurement in the basis $\{(\ket{H}\pm\ket{V})/\sqrt{2}\}$ projects the three-atom state onto the GHZ state up to local Pauli gates, $(\ket{+++}_a \pm \ket{---}_a)/\sqrt{2}$.

For realistic cavity systems, however, the reflection coefficients for the atom $\ket{0}_a$ and $\ket{1}_a$ are respectively given by~\cite{Cohen2018, Utsugi2025}
\begin{equation}
    \begin{aligned}
        r_0(\Delta) =& \frac{-\kappa_\text{ex}+ \kappa_\text{in} - i\Delta}{\kappa_\text{ex} + \kappa_\text{in}-i\Delta}, \\
        r_1(\Delta) =& \frac{(-\kappa_\text{ex} + \kappa_\text{in}-i\Delta)(\gamma -i\Delta) + g^2}{(\kappa_\text{ex} + \kappa_\text{in}-i\Delta)(\gamma -i\Delta) + g^2},
    \end{aligned}
\end{equation}
where $\Delta$ represents the detuning of the photon from the cavity.
To mitigate the effect of finite optical loss and pulse delay in photon reflection, we apply a calibrated loss and delay line to maximize the overlap of the photonic envelope between different polarized photons (see the details in Ref.~\cite{2507.01229}).
Here, the amplitudes $r_{j}^\text{opt}$ and delays $\tau_j$ are independently determined by each cavity parameters~\cite{2507.01229},
\begin{equation}
    r_{j}^\text{opt} = 1- \frac{2}{1+\sqrt{1+2C_{\text{in},j}}}, \quad \tau_j = \frac{2\kappa_{\text{ex},j}}{\kappa_{\text{ex},j}^2-\kappa_{\text{in},j}^2}.
\end{equation}
In the following description of a photon-atom state, we omit the term for photon loss because it is finally excluded by detector click events.
We present the (unnormalized) photon-atom state just in front of the quarter-wave plate as
\begin{widetext}
    \begin{equation}
        \ket{\psi} = \frac{1}{\sqrt{2}} \int \dd{\Delta} f(\Delta) \ab[\hat{a}_0^\dagger(\Delta) \bar{r}e^{i\bar{\tau}\Delta}\ket{+++}_a + \hat{a}_1^\dagger(\Delta) \bigotimes_{j\in\{1,2,3\}} \frac{r_{0,j}(\Delta)\ket{0}_{a,j} + r_{1,j}(\Delta)\ket{1}_{a,j}}{\sqrt{2}}]\ket{\bm{0}}_p,
    \end{equation}
\end{widetext}
where $\hat{a}_{0(1)}(\Delta)$ represents the annihilation operator for the $H(V)$-polarized photon at detuning $\Delta$, $f(\Delta)$ represents the spectrum of the prepared photon, and $\ket{\bm{0}}_p$ represents the vacuum state in the propagating modes.
Here, we have defined $\bar{r} = \Pi_{j\in\{1,2,3\}} r_{j}^\text{opt}$ and $\bar{\tau} = \sum_{j\in\{1,2,3\}} \tau_j$, and the coefficient $\bar{r} e^{i\bar{\tau} \Delta}$ represents the total calibration loss and delay.
For a detector having a flat frequency response, the positive operator-valued measure (POVM) of the photon detection is given by $\{\Pi_\pm\}$ with
\begin{equation}
    \Pi_{\pm} = \int \dd{\Delta} \frac{\hat{a}_0^\dagger(\Delta)\pm\hat{a}_1^\dagger(\Delta)}{\sqrt{2}}\ketbra{\bm{0}}[_p]{\bm{0}} \frac{\hat{a}_0(\Delta)\pm\hat{a}_1(\Delta)}{\sqrt{2}}.
\end{equation}
From the relation
\begin{equation}
    {}_p\bra{\bm{0}}\frac{\hat{a}_0(\Delta)\pm\hat{a}_1(\Delta)}{\sqrt{2}}\ket{\psi} = \frac{f(\Delta)}{\sqrt{2}}\ket{\Upsilon_{\pm}(\Delta)}_a,
\end{equation}
and 
\begin{equation}
    \begin{aligned}
        \ket{\Upsilon_{\pm}(\Delta)}_a = \frac{1}{\sqrt{2}}&\Bigg[\bar{r}e^{i\bar{\tau}\Delta}\ket{+++}_a \\
        &\pm \bigotimes_{j} \frac{r_{0,j}(\Delta)\ket{0}_{a,j} + r_{1,j}(\Delta)\ket{1}_{a,j}}{\sqrt{2}}\Bigg],
    \end{aligned}
\end{equation}
the postmeasurement state is given by
\begin{equation}
    \begin{aligned}
        \rho_\pm =& \frac{\Tr_p[\Pi_{\pm} \ketbra{\psi}{\psi}]}{\Tr[\Pi_{\pm} \ketbra{\psi}{\psi}]} \\
        =& \frac{1}{p_{\pm}} \int \dd{\Delta}\frac{\vab{f(\Delta)}^2}{2} \ketbra{\Upsilon_{\pm}(\Delta)}[_a]{\Upsilon_{\pm}(\Delta)},
    \end{aligned}
\end{equation}
where $p_\pm$ represents the detection probability:
\begin{equation}
    p_\pm = \Tr[\Pi_{\pm} \ketbra{\psi}{\psi}] = \int \dd{\Delta}\frac{\vab{f(\Delta)}^2}{2} \braket{\Upsilon_{\pm}(\Delta)}{\Upsilon_{\pm}(\Delta)}.
\end{equation}
The average fidelity is then given by
\begin{equation}
    F_\text{GHZ} = \frac{p_{+} \bra{\text{GHZ}_-}\rho_+\ket{\text{GHZ}_-} + p_{-} \bra{\text{GHZ}_+}\rho_-\ket{\text{GHZ}_+}}{p_{+} + p_{-}},
\end{equation}
and the total success probability is $p_{+} + p_{-}$.

\end{document}